\documentclass[a4paper, amsfonts, amssymb, amsmath, reprint, showkeys,nofootinbib,twoside,unsortedaddress,superscriptaddress,floatfix]{revtex4-1}

\usepackage[english]{babel}
\usepackage[utf8]{inputenc}
\usepackage[colorinlistoftodos, color=green!40, prependcaption]{todonotes}
\usepackage{aas_macros}

\usepackage{amsthm}
\usepackage{mathtools}
\usepackage{physics}
\usepackage{xcolor}
\usepackage{graphicx}
\usepackage[left=23mm,right=13mm,top=35mm,columnsep=15pt]{geometry}
\usepackage{adjustbox}
\usepackage{placeins}
\usepackage[T1]{fontenc}
\usepackage{lipsum}
\usepackage{csquotes}
\usepackage{booktabs}
\usepackage{array}

\usepackage[pdftex, pdftitle={Article}, pdfauthor={Author}]{hyperref} 
\begin{document}
\title{Relevance of photon-photon dispersion within the jet for blazar axionlike particle searches}


\author{James Davies}
\email[Email address: ]{james.davies2@physics.ox.ac.uk}
\affiliation{University of Oxford, Department of Physics, Oxford, UK}

\author{Manuel Meyer}
\email[Email address: ]{manuel.meyer@desy.de}
\affiliation{Institute for Experimental Physics, University of Hamburg, Luruper Chaussee 149, D-22761 Hamburg, Germany}

\author{Garret Cotter}
\email[Email address: ]{garret.cotter@physics.ox.ac.uk}
\affiliation{University of Oxford, Department of Physics, Oxford, UK}

\date{\today} 

\begin{abstract}
Axionlike particles (ALPs) could mix with photons in the presence of astrophysical magnetic fields. Searching for this effect in gamma-ray observations of blazars has provided some of the strongest constraints on ALP parameter space so far. Previously, photon-photon dispersion of gamma-rays off of the CMB has been shown to be important for these calculations, and is universally included in ALP-photon mixing models. Here, we assess the effects of dispersion off of other photon fields within the blazar (produced by the accretion disk, the broad line region, the dust torus, starlight, and the synchrotron field) by modelling the jet and fields of the flat spectrum radio quasar 3C454.3 and propagating ALPs through the model both with and without the full dispersion calculation. We find that the full dispersion calculation can strongly affect the mixing, particularly at energies above 100 GeV -- often reducing the ALP-photon conversion probability. This could have implications for future searches planned with, e.g., the Cherenkov Telescope Array, particularly those looking for a reduced opacity of the universe at the highest energies.
\end{abstract}


\maketitle

\section{Introduction} \label{sec:intro}
The axion is an undiscovered particle beyond the Standard Model (BSM) of particle physics, theorized in order to solve the strong CP problem \cite{Weinberg_78, Wilczek_78, Peccei_06}. It is a very light (sub-eV) pseudoscalar particle that couples to photons in an external magnetic field. Axionlike particles (ALPs) are similar particles, also coupling to photons, that commonly arise in string theories, or as pseudo-Nambu-Goldstone bosons in other BSM theories \cite{Ringwald_2014, Irastorza_2018}. While no longer necessarily solving the CP problem, ALPs could, for certain values of mass and coupling, make up all or some of dark matter \cite{Arias_2012}.\par
Axionlike particles are therefore important candidates for both direct and indirect searches. Many searches utilise the fact that, in the presence of an external magnetic field, ALPs will oscillate into photons and vice-versa in a way analogous to neutrino oscillations \cite{Raffelt_1988}. The oscillation length of this conversion depends on the ALP mass, $m_a$, its coupling to photons, $g_{a\gamma}$, as well as the properties of the external medium -- most notably the magnetic field strength, $B$ (required for mixing in order to conserve spin) \cite{deAngelis_Galanti_Roncadelli_Rev_2011}. This would lead to many potentially observable effects (see, e.g., \cite{CAST_2017,Payez_2015,Battaglieri_2017}). \par
Astrophysical gamma-ray observations have provided some of the best limits in ALP ($m_a$, $g_{a\gamma}$) parameter space (e.g., \cite{Payez_2015}). The focus of this work is on ALP searches that look for irregularities in blazar gamma-ray spectra. Blazars are jetted active galactic nulcei (AGN -- a supermassive black hole producing bright radiation as it accretes surrounding matter) which have their relativistic jets pointed towards us (within a few degrees). These jets of plasma produce non-thermal emission across the whole electromagnetic spectrum by synchrotron or inverse-Compton\footnote{It could also be possible for hadronic models to produce the high energy emission from, e.g., proton-synchrotron, proton-pion production, and photo-pion production (e.g. \cite{Petropoulou_2012, Mucke_2003}) though they often require super-Eddington powered jets \cite{Zdziarski_2015}.} emission, the brightness of which is strongly enhanced by relativistic effects \cite{Blandford_Rev_2019}. In particular, they constitute nearly all of the known extragalactic gamma-ray sources \cite{Fermi_4FGL_2020}. Irregularities in blazar gamma-ray spectra could be caused by ALP-photon mixing in astrophysical magnetic fields along the line of sight to the source (e.g., \cite{Fermi_2016, HESS_2013, CTA_Gpropa_2021}). An energy dependent oscillation length within certain magnetic field environments could lead to oscillatory features in the otherwise-smooth blazar gamma-ray spectra. Even without oscillatory features, ALP-photon mixing could reduce the apparent opacity of the universe at the highest energies as photons temporarily converted to ALPs would not suffer from pair-production off of background fields (e.g.,  \cite{Sanchez-Conde_2009,DeAngelis_2007,DeAngelis_2009,deAngelis_Galanti_Roncadelli_Rev_2011,Meyer_Conrad_2014}). The magnetic field environments used for these searches have been the ubiquitous intergalactic magnetic field (IGMF, e.g, \cite{Montanino_2017,Galanti_Smoothed_2018}), the galactic magnetic field (GMF, e.g., \cite{Majumdar_2018, Carenza_2021}), and source-dependent galaxy cluster fields (CMF, e.g., the Perseus Cluster field surrounding NGC 1275 used in \cite{Fermi_2016, Reynolds_2020}). Also, the magnetic field within the source itself has been shown to be a promising environment for extending ALP constraints (e.g., \cite{Hochmuth_2007, Tavecchio_2015, Davies_2021}).\par
Ref. \cite{Dobrynina_2015} showed that, because these searches depend on slight differences between propagation in the ALP and photon states, the very small effect of photon-photon dispersion off of the CMB can make a discernible difference and should be included in the calculations. Photon-photon dispersion adds to the dispersive part, $\chi$, of the refractive index of the propagation medium, $n=1+\chi+i\kappa$ ($\kappa$ is the absorptive part). Here, we investigate the effects of dispersion off of other fields that could be found within the source. Blazars can be broadly classified into flat spectrum radio quasars (FSRQs) and BL Lacs depending on the strength or weakness respectively of broad emission lines in their optical spectra (for reviews, see, e.g., \cite{Blandford_Rev_2019, Hovatta_Lindfors_Rev_2019}). These broad lines imply, for FSRQs, the presence of rapidly-moving clouds of gas close to the central black hole -- the broad line region (BLR) -- which are reprocessing accretion disk photons. Infrared observations also imply the presence of a dusty molecular torus beyond the BLR, reprocessing disk photons at a lower energy. For FSRQs then, background photons from the accretion disk, the BLR, the dust torus, as well as the synchrotron photons produced in the jet itself can be expected within the source. Also, unrelated to the AGN, starlight from the host galaxy should be present. To investigate the importance of dispersion off of these fields, we model the jet (Sec. \ref{subsec:jet}; as in Ref. \cite{Davies_2021}) and fields (Secs. \ref{sec:fields} to \ref{subsec:u_tot}; following Ref. \cite{Finke_2016}) of the FSRQ 3C454.3, and simulate broad-band spectral energy distributions (SEDs) for both a flaring and a quiescent state, checking them against observations (Sec. \ref{subsec:synch}). We then propagate ALP-photon beams through our model both with and without the full dispersion calculation and compare the photon-survival probabilities ($P_{\gamma\gamma}$s) between the two (Sec. \ref{sec:chi}).

\section{Modelling 3C454.3} \label{sec:modelling}
\begin{figure}
  \centering
  \includegraphics[width=0.48\textwidth]{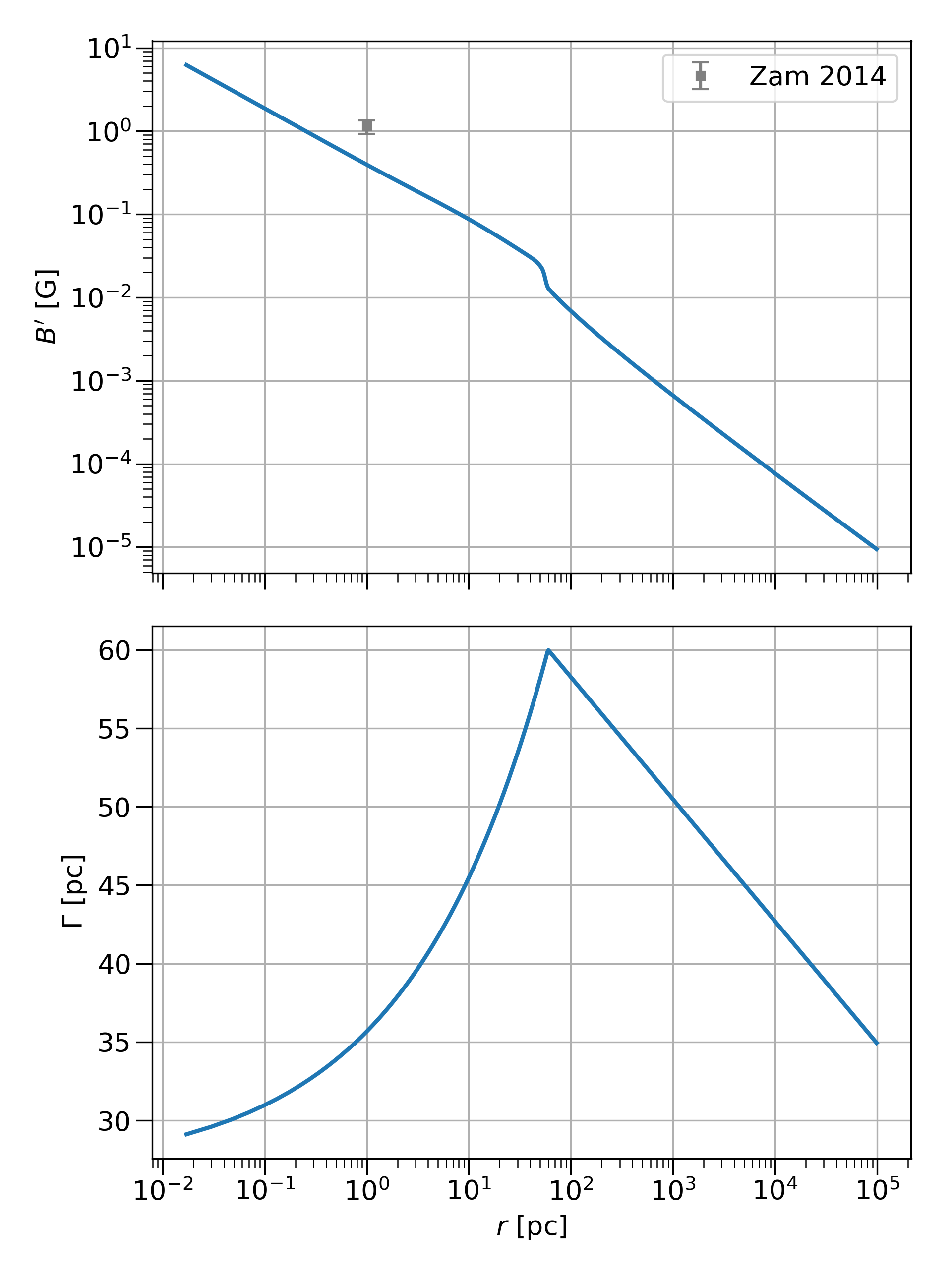}
    \caption{(Top) Total magnetic field $B'$ vs $r$ showing the transition region from the highly magnetised parabolic base to the conical jet in rough equipartition. Estimate from Ref. \cite{Zamaninasab_2014} plotted for comparison.
  (Bottom) Bulk Lorentz factor $\Gamma$ vs $r$. $\Gamma$ increases, peaks at 60 at the transition region, then decelerates logarithmically down the jet.}
  \label{fig:B}
  \label{fig:G}
\end{figure}
To assess the possible effects of dispersion within the jet, we need a model of the jet to propagate ALPs through, and a model of the relevant background photon fields. We choose 3C454.3, a powerful FSRQ at redshift $z=0.859$ that is bright in gamma rays and has been observed in both flaring and quiescent states.
\subsection{Jet} \label{subsec:jet}
 We model the jet within the Potter \& Cotter framework (PC, see \cite{Potter_Cotter_NC_2015}) as discussed in the context of ALP-photon mixing in Ref. \cite{Davies_2021}\footnote{And available within the \texttt{gammaALPs} \texttt{PYTHON} package, hosted on GitHub (\url{https://github.com/me-manu/gammaALPs}) and archived on Zenodo \cite{gammaALPs_2021}. See Ref. \cite{Meyer_ICRC_2021} for an overview.}. The structure of this jet is an accelerating, parabolic, magnetically dominated jet base which transitions to a decelerating conical ballistic jet in rough equipartion at around $10^5 r_g$. This parabolic shape is seen in observations of M87, and a highly magnetised base is consistent with magnetic jet launching mechanisms, e.g. the Blandford-Znajek mechanism \cite{Blandford_Znajek_1977}. Within the parabolic base, the overall field strength drops as $r^{-a}$ where $a\sim0.58$ is the parabolic index, and in the conical jet $B$ drops roughly as $r^{-1}$, consistent with VLBI observations \cite{Pudritz_Hardcastle_Gabuzda_LaunchTerm_2012}. The PC model is able to produce broadband steady-state SEDs for many blazars which fit observations well from radio to gamma rays \cite{Potter_Cotter_NC_2015}. Within the PC model, the steady-state gamma-ray emission is strongly dominated by the transition region ($r_{\mathrm{vhe}}=59.8$ pc for 3C454.3). A large portion of the synchrotron emission is also produced here, though synchrotron emission from the rest of the jet is also important. \par Figure \ref{fig:B} shows how the overall magnetic field strength varies along our jet. The field at the transition region is, $B(r_{\mathrm{vhe}}) = 1.3\times10^{-2}$ G, which means the field at $1$ pc is slightly lower than the estimate from Ref. \cite{Zamaninasab_2014}, though they use a conical jet model at all distances. Figure \ref{fig:G} also shows the bulk Lorentz factor, $\Gamma$ along the jet. As the jet accelerates in the parabolic base, the bulk Lorentz factor increases ($\Gamma \propto r^{1/2}$) up to a maximum of 60. Beyond the transition region, as the jet decelerates, it decreases ($\Gamma\propto \log(r)$), until the end of the jet, $r_{jet}=100$ kpc. The jet radius at the transition region is about $R=0.7$ pc, and the electron density is (from equipartition considerations) $n_{e}=4.71$ cm$^{-3}$. The electron density varies as $n_{e}\propto R^{-2}$ as usual to conserve particle number (though notably this means $n_{e} \neq r^{-2}$ in the parabolic base).

\subsection{Photon Fields} \label{sec:fields}
Far from pair production energies, a background electromagnetic field of energy density $u$ will lead to dispersion $\chi \propto u$ (see Eq. (1) in Ref. \cite{Dobrynina_2015}). At $z=0$, this low energy dispersion off the isotropic CMB field is $\chi_{CMB}=5.11\times10^{-43}$. This term and the analogous dispersion off the $B$ field itself, are usually included in ALP works.\par
More generally, from equation (8) in Ref. \cite{Dobrynina_2015}, the dispersion of a photon with energy $\omega$ off of a photon field with differential energy density $u(\epsilon, \Omega$) at energies $\epsilon$ (related to frequency by $\epsilon=h\nu$) is given by:

\begin{equation}\label{eq:chi}
\begin{split}
    \chi(\omega)=&\frac{44\alpha^2}{135 m_e^4}\int^{2\pi}_{0}d\phi \int_{-1}^{1}d\mu \\
    & \times \int_0^{\infty}d\epsilon u(\epsilon,\Omega) \frac{3}{4}(1-\mu)^2\text{g}_0\Big(\frac{\omega}{\omega_0}\Big)
\end{split}
\end{equation}
with $\mu=\cos{\theta}$ where $\theta$ is the angle of the background photon with respect to the dispersed photon. $\omega_0$ is the threshhold energy for pair-production, $\omega_0=2m_e^2/\epsilon(1-\mu)$. The function, $\text{g}_0$ is given in equation (9) of Ref. \cite{Dobrynina_2015} and describes the dispersion energy dependence around pair-production energies. This more general form of $\chi$ is also valid around pair-production energies and includes an angular dependence of the background field.\par
Of course, above pair-production energies, a background photon field will also lead to the absorption of gamma rays. The absorption rate for this process is given by,
\begin{equation}\label{eq:gamma}
\begin{split}
    \Gamma_{\gamma\gamma}(\omega)=&\int^{2\pi}_{0}d\phi \int_{-1}^{1}d\mu \\
    & \times \int_0^{\infty}d\epsilon (1-\mu) n(\epsilon,\Omega) \sigma_{\gamma\gamma}(\omega,\epsilon,\mu),
\end{split}
\end{equation}
where $n(\epsilon,\Omega)=u(\epsilon,\Omega)/\epsilon$ is the differential number density of the background field and $\sigma_{\gamma\gamma}=(\pi\alpha^2/2m_e^2)f(\omega/\omega_0)$ is the total $\gamma\gamma \to e^+ e^-$ cross section. The function $f(\omega/\omega_0)$ is given in equation (7) of Ref. \cite{Dobrynina_2015}.
It does not make sense to consider the effects of dispersion off of background fields at energies where absorption from these same fields is significant. Also, $\Gamma_{\gamma\gamma}$ must be included in the ALP-photon mixing equations because ALPs are not absorbed by pair-production, but photons are. Therefore, an investigation into the effects of dispersion on ALP-photon mixing should also include absorption.\par
To calculate the dispersion and absorption using Eqs. \eqref{eq:chi} and \eqref{eq:gamma}, we require the differential energy densities, $u(\epsilon, \Omega)$ of the relevant fields. These are the CMB, the extragalactic background light (EBL), starlight (SL) from the host galaxy, the AGN fields (accretion disk, broad line region, dust torus), and the synchrotron field within the jet.

\subsection{CMB}
The CMB field at a redshift $z$ is an isotropic black-body radiation field with temperature, $T=T_0(1+z)$, where $T_0=2.726$ K. The spectral energy density of the CMB is the black-body spectral radiance, $B_\nu$, divided by $c$,

\begin{equation}\label{eq:u_cmb}
    u(\nu, \Omega) = \frac{B_\nu}{c} = \frac{2h\nu^3}{c^3}\frac{1}{e^{\frac{h\nu}{k_B T}}-1},
\end{equation}
which can be converted into the differential energy density by dividing by $h$: $u(\epsilon, \Omega) = u(\nu, \Omega)/h$. Note that even in the low energy case, the redshift dependence of the CMB temperature (and therefore energy density) means that the term included in ALP calculations should be $\chi_{CMB}(1+z)^4$ instead of simply $\chi_{CMB}$.

\subsection{EBL and starlight}\label{subsec:ebl_sl}
The EBL is the total starlight emitted and reprocessed throughout the history of the universe. Most of the emission lies in the ultraviolet to infrared range, with direct starlight emission at the high-frequency end and reprocessed emission at lower frequencies. We use the EBL model from Ref. \cite{Dominguez_EBL_2011}. The \texttt{ebltable} \texttt{PYTHON} package\footnote{Available at : \url{https://github.com/me-manu/ebltable}} allows the extraction of the EBL photon density at a given redshift, $dn/d\epsilon$, which gives the differential energy density, $u(\epsilon,\Omega)=(\epsilon/4\pi)\times dn/d\epsilon$.\par
Most of the starlight seen within the jet, however, is produced from the host galaxy itself. We follow Ref. \cite{Potter_Cotter_2_2013} in modelling the host galaxy as a giant elliptical, as would be expected for a high-powered FSRQ such as 3C454.3. The observed luminosity density of large elliptical galaxies as a function of distance from the galaxy center stays quite flat within the core ($\rho_L\propto r^a$ with $-1.3\leq a\leq -1$) and then decreases more rapidly at larger distances. At $r_{30}=30$ pc, typical luminosity densities of large ellipticals are between 2 and 10 $L_{\odot}$ pc$^{-3}$; we take $\rho_{30}=5L_{\odot}$ as in Ref. \cite{Potter_Cotter_2_2013}. We use $r_{core}=20$ kpc, which seems to be typical for blazar hosts (see, e.g., Refs. \cite{Olguin-Iglesias_2016,Nilsson_2003}). Specifically, for the luminosity densities, we use,

\begin{equation}\label{eq:rho_L}
    \rho_L (r) =
    \begin{dcases}
    \rho_{30} \big(\frac{r_{30}}{r}\big)^{1.2},& \text{if } 1 \text{pc}\leq x\leq r_{core}\\
    \rho_L(r_{core})\big(\frac{r_{core}}{r}\big)^4,              &\text{if } x > r_{core}
\end{dcases}
\end{equation}
with a constant luminosity density for $r<1$ pc and a maximum galaxy radius of 100 kpc. For Eq. \ref{eq:chi}, we also require the energy dependence of the field. We use a black-body spectrum, $B_\nu^{T_K}$, at a temperature of $T_K =$ 5000 K to approximate the emission of a typical K-type star, which should dominate the emission of an elliptical galaxy. We calculate $u(\epsilon,\Omega)$ as in Ref. \cite{Finke_2016}. Each point in the galaxy emits isotropically, with emissivity
\begin{equation}\label{eq:emissivity}
    j(\epsilon,\Omega_g;R_{\mathrm{gal}})=\rho_L(R_{\mathrm{gal}})\frac{B_\nu^{T_K}}{\int d\Omega_{\mathrm{gal}}\int d\epsilon B_\nu^{T_K}}
\end{equation}
where $R_{\mathrm{gal}}$ is the radial distance from the centre of the galaxy and $\Omega_{\mathrm{gal}}$ is the solid angle measured in the galaxy frame. A point in the galaxy at angle $\mu_{\mathrm{gal}}=\cos(\theta_{\mathrm{gal}})$ above the plane perpendicular to the jet and a distance $R_{\mathrm{gal}}$ from the galaxy centre will be a distance $x^2=R_{\mathrm{gal}}^2+r^2-2rR_{\mathrm{gal}}\mu_{\mathrm{gal}}$ away from a point $r$ along the jet. From $r$, this point will be at an angle
\begin{equation}\label{eq:mu_r}
    \mu_{r}^2=1-\Big(\frac{R_{\mathrm{gal}}}{x}\Big)^2(1-\mu_{\mathrm{gal}}^2).
\end{equation}
Overall then, integrating over the galaxy, the starlight field observed at $r$ will be
\begin{equation}\label{u_star}
\begin{split}
    u(\epsilon,\Omega;r)=&\frac{1}{4\pi c}\int_0^{2\pi} d\phi_{\mathrm{gal}}\int_{-1}^{1}d\mu_{\mathrm{gal}} \int_{0}^{\infty} d R_{\mathrm{gal}} \\
    & \times \Big(\frac{R_{\mathrm{gal}}}{x}\Big)^2 j(\epsilon,\Omega_{\mathrm{gal}};R_{\mathrm{gal}})\\
    & \times \delta(\phi - \phi_{\mathrm{gal}})\delta(\mu - \mu_r).
\end{split}
\end{equation}

\begin{table}
 \caption{Parameters used for the AGN Fields.}
  \centering
  \begin{tabular}{m{0.1\textwidth} m{0.18\textwidth}}
    \toprule\toprule
    \multicolumn{2}{c}{Disk} \\
    \midrule
    $L_{disk}$ & $2\times10^{46}$ erg s$^{-1}$ \\
    $M_{BH}$ & $1.2\times10^{9} M_{\odot}$ \\
    $R_{in}$ & $6 R_g$ \\
    $R_{out}$ & $200 R_g$ \\
    $\eta$ & 1/12 \\
    \midrule\midrule
    \multicolumn{2}{c}{Broad Line Region} \\
    \midrule
    $L_{H\beta}$ & $4.18\times10^{43}$ erg s$^{-1}$ \\
    $R_{H\beta}$ & $4.3 \times 10^{17}$ cm \\
    \midrule\midrule
    \multicolumn{2}{c}{Torus} \\
    \midrule
    $\Theta$ & $10^3$ K \\
    $\xi_{dt}$ & 0.1 \\
    $\zeta$ & 1 \\
    $R_1$ & $1.6\times10^{19}$ cm \\
    $R_2$ & $1.6\times10^{20}$ cm \\
    $b/a$ & 0.527 ($f_c=0.6$) \\
    \bottomrule\bottomrule
  \end{tabular}
  \label{tab:fields}
\end{table}

\subsection{AGN fields} \label{subsec:agn}
The models of the AGN photon fields are based on those from Ref. \cite{Finke_2016}, which uses them in the context of Compton-scattering calculations. The sources of the disk and BLR fields are modelled as thin, i.e., they only extend radially from the black hole and do not extend vertically in the direction of the jet. This radial coordinate is denoted by $R$, and the distance from a point $R$ to a point $r$ in the jet is given by $x=\sqrt{R^2 + r^2}$. Therefore, in the galaxy frame, a photon emitted at $R$ will have an incident angle $\mu=\cos{\theta}=r/x$ at $r$ (see Figure 9. in Ref. \cite{Finke_2016}). For the torus, we extend the flat model of Ref. \cite{Finke_2016} to include an elliptical cross-section.

\subsubsection{Disk} \label{subsubsec:disk}
The accretion disk is modelled as a flat Shakura-Sunyaev disk \cite{Shakura_Sunyaev_1973}, but with a delta approximation meaning each radius ($R$) emits isotropically but at only one frequency, $\epsilon_0 m_e c^2$. The differential energy density can be written,

\begin{equation}\label{eq:u_disk}
    u(\epsilon, \Omega; r) = \frac{3}{(4 \pi)^2 c}\frac{L_{disk} R_g}{\eta R^3 \mu} \varphi(R) \delta(\epsilon -\epsilon_0(R)),
\end{equation}
with
\begin{equation}\label{eq:phi}
    \varphi(R) = \sqrt{1-\frac{R_{in}}{R}},
\end{equation}
and
\begin{equation}\label{eq:e0}
    \epsilon_0(R) = 2.7\times10^{-4}\Big(\frac{l_{Edd}}{M_8 \eta}\Big)^\frac{1}{4}\Big(\frac{R}{R_g}\Big)^{-\frac{3}{4}},
\end{equation}
where $L_{disk}$ is the disk luminosity, $\eta$ is the accretion efficiency, the black hole mass is $M_{BH}=10^8 M_8 M_\odot$ and its gravitational radius is $R_g=G M_{BH}/c^2$. $R_{in}$ is the inner radius of the accretion disk ($6R_g$ for a non-rotating Schwarzschild black hole, $R_g$ or $9 R_g$ for a disk rotating in either the same or opposite direction as a spinning black hole respectively \cite{Tsupko_2016}).

\subsubsection{BLR} \label{subsubsec:blr}
The BLR is similarly approximated as a series of rings, where each ring (at a given $R_{li}$) corresponds to a line in the BLR spectrum and emits isotropically at only one energy, $\epsilon_{li}$. GRAVITY observations of BLRs support a ringlike BLR model as opposed to a shell-like one \cite{Gravity_BLR_2018}. Also, the optical depths of the concentric-ring model of Ref. \cite{Finke_2016} agree with those found by a detailed modelling of the BLR in Ref. \cite{Abolmasov_Poutanen_2017}, as discussed in Ref. \cite{Meyer_Scargle_Blandford_2019}. We use all the lines in the appendix of Ref. \cite{Finke_2016}. For each line,

\begin{equation}\label{eq:u_blr}
    u(\epsilon, \Omega; r) = \frac{\xi_{li}L_{disk}}{(4\pi)^2 c x^2}\delta(\mu - r/x)\delta(\epsilon - \epsilon_{li}),
\end{equation}
where $\xi_{li}$ is the fraction of the disk luminosity re-emitted by the line, $L_{li}=\xi_{li}L_{disk}$. Reverberation mapping indicates that the ratios of line luminosities and line radii between lines within a source remain roughly constant between sources. This means that estimates of the luminosities, $L_{li}$ (or equivalently $\xi_{li}$) and radii $R_{li}$ (needed for $x$) of the lines can be obtained from an estimate of $L_{H\beta}$ and $R_{H\beta}$ for 3C454.3. Finke does this in the appendix of \cite{Finke_2016}, the values are listed in Table \ref{tab:fields}.

\subsubsection{Torus} \label{subsubsec:dt}
For the torus, we extend the flat disk model of Ref. \cite{Finke_2016} to account a non-zero torus height. In particular, we include an elliptical torus cross-section. This means the ratio of the semi-minor to semi-major axes ($b/a$) can be adjusted to produce a particular covering fraction (the fraction of the sky obscured by the torus, from the position of the black hole), typically around $f_c\sim0.6$ (see, e.g., Refs. \cite{Calderone_2012,Galanti_2020}). Because the geometry of this model is a bit more involved, we leave a detailed discussion to the Appendix \ref{app:app}.\par
Aside from the geometry, the other features of the model are taken from Ref. \cite{Finke_2016}. The torus radiates a fraction of the disk luminosity, $\xi_{dt}$, at a single energy, $\epsilon_{dt}=2.7\Theta$, which depends on the temperature of the torus, $\Theta$. The fraction of the emission emitted at each $R$ drops $\propto R^{-\zeta}$ to account for the decrease in cloud number density through the torus, from the inner radius ($R_1$) to the outer radius ($R_2$). Table \ref{tab:fields} shows the values of the field parameters used for 3C454.3.




\subsection{Transformations}\label{subsec:transformations}
In this model, the AGN fields all emit isotropically in the galaxy frame and each radius, $R$, from the black hole emits at only one energy. This means the photon fields in the co-moving jet frame are anisotropic and non-thermal, i.e., not described by a black-body spectrum. Equation \eqref{eq:chi} requires the energy densities and angles to be measured in the same frame as the dispersed photon energy, $\omega$. We choose the co-moving jet frame (denoted by a prime) as opposed to the galaxy frame as this will simplify the synchrotron calculations (Sec. \ref{subsec:synch}) and because this is the frame the rest of the ALP calculations are done in. It is worth stressing that, because the bulk Lorentz factor, $\Gamma$, changes along the jet (see Fig. \ref{fig:G}), the jet frame is a local frame--i.e, it will depend on $r$. The energies, energy-densities, and incident angles of the background photon fields therefore need to be transformed into the jet frame at each $r$. This is done through the Doppler factor, $\delta$ which will depend on $r$ through $\Gamma$:

\begin{equation}\label{eq:delta}
    \delta(\mu;r) = [\Gamma(1 - \beta\mu)]^{-1}.
\end{equation}

Then,

\begin{equation}\label{eq:e}
    \epsilon' = \epsilon \delta^{-1},
\end{equation}

\begin{equation}\label{eq:u}
    u'(\epsilon', \Omega') = u(\epsilon, \Omega)\delta^{-3},
\end{equation}
as $u(\epsilon, \Omega)/\epsilon^3$ is invariant \cite{Dermer_2002}. The angles transform as

\begin{equation}\label{eq:mu}
    \mu' = \frac{\mu - \beta}{1 - \beta\mu},
\end{equation}
where $\beta = \sqrt{1 - 1/\Gamma^2}$.
Using these transformations, $d\mu'=\delta(\mu)^2 d\mu$, $d\epsilon'=\delta(\mu)^{-1}d\epsilon$ and $d\phi'=d\phi$. Then from Eq. \eqref{eq:chi}, inserting $u$ for $u'$ from Eq. \eqref{eq:u}, $\chi$ for the AGN fields can be found by,

\begin{equation}\label{eq:chi_prime}
\begin{split}
    \chi(\omega';r)=&\frac{44\alpha^2}{135 m_e^4}2\pi \int_{\mu_{min}}^{\mu_{max}}d\mu\\
    & \times \int_0^{\infty}d\epsilon \frac{u(\epsilon,\Omega)}{\delta(\mu)^2} \frac{3}{4}(1-\mu')^2\text{g}_0\Big(\frac{\omega'}{\omega_0'}\Big).
\end{split}
\end{equation}
Note that, even though the jet is receding from the fields by the black hole, all photons with $\mu < \beta$ ($0.9993<\beta<0.9999$ in our jet) will have $\mu'<0$ and so will be seen in the head-on hemisphere in the jet frame. This means that, counterintuitively, the $(1-\mu')^2$ term in Eq. \eqref{eq:chi_prime} can actually increase the dispersion off of the AGN fields as opposed to suppressing it. Similarly, most of the CMB photons are viewed close to head-on, which enhances the dispersion off of the CMB. This makes sense by comparison to the well-known Doppler boosting of blazar synchrotron radiation. A synchrotron radiation field that is seen as isotropic in the jet frame is seen as close to co-linear to the jet in the galaxy frame. Therefore, doing the reverse transformation, photons seen as close to co-linear in the galaxy frame can be seen as isotropic (i.e. more head-on) in the jet frame.\par
The pair-production absorption rate can be similarly transformed:
\begin{equation}\label{eq:gamma_prime}
\begin{split}
    \Gamma_{\gamma\gamma}(\omega';r)=&\int^{2\pi}_{0}d\phi \int_{-1}^{1}d\mu \\
    & \times \int_0^{\infty}d\epsilon (1-\mu') \frac{n(\epsilon,\Omega)}{\delta(\mu)} \sigma_{\gamma\gamma}(\omega',\epsilon',\mu').
\end{split}
\end{equation}
Regarding absorption, it is useful to calculate the optical depth of the background photon fields in the observed frame. This can be done by accounting for the redshift transformation and the ($r$-dependent) Lorentz transformation from the co-moving jet frame. Because the jet frame is a local frame, a given observed energy, $\omega_{obs}$, will correspond to different jet-frame energies along the jet. Assuming we look directly down the jet, $\omega'(r)=\omega_{obs}(1+z)/2\Gamma(r)$. $\Gamma_{\gamma\gamma}$ itself also has to be transformed (not being dimensionless): $\Gamma_{\gamma\gamma}^{obs}=2\Gamma(r)\Gamma_{\gamma\gamma}$. The optical depth of the fields for a photon propagating from $r_0$ is then,
\begin{equation}\label{eq:tau_obs}
    \tau(\omega_{obs};r_0)=2\int^{r_{jet}}_{r_0} dr \Gamma(r) \Gamma_{\gamma\gamma}(\omega';r).
\end{equation}

\begin{figure}
  \centering
    \includegraphics[width=0.48\textwidth]{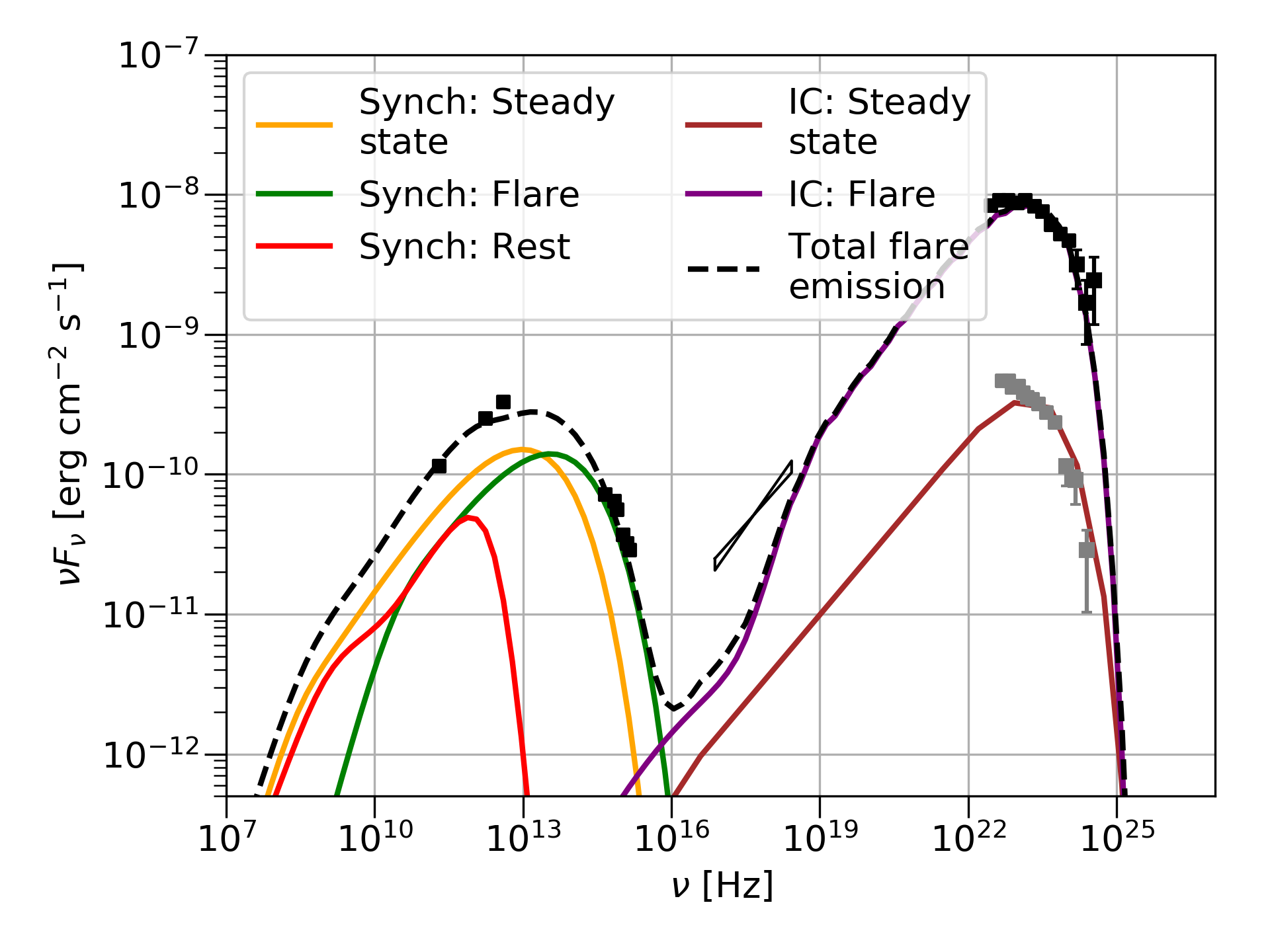}
  \caption{SED from our jet model, showing synchrotron and inverse-Compton emission from the flare and steady state regions, and the synchrotron emission from the rest of the jet. Data from Ref. \cite{Cerruti_2013}. Black points show data taken during a flare, grey points show steady-state emission.}
  \label{fig:sed}
\end{figure}
\begin{table}
 \caption{Parameters used for the blobs down the jet. The steady-state and flaring gamma-ray emission is produced from blobs at $r_{\mathrm{vhe}}$ and $r_{\mathrm{em}}$ respectively.}
  \centering
  \begin{tabular}{m{0.1\textwidth} m{0.12\textwidth} m{0.12\textwidth}m{0.1\textwidth}}
    \toprule\toprule
    Parameter & $r_{\mathrm{vhe}}$ & $r_{\mathrm{em}}$  &
    rest of jet\\
    \midrule
    $r$ (pc) & 59.8 & 0.1 & -  \\
    $B$ (G) & 0.013 & 1.9 & -\\
    $n_e$ cm$^{-3}$ & 4.71  & 7.9$\times$10$^3$ & - \\
    $E_c$ (MeV) & 1.3$\times10^3$ & 250 & 1.3 \\
    $\beta$ & 1.95 & 2 & 2  \\
    \bottomrule\bottomrule
  \end{tabular}
  \label{tab:zones}
\end{table}
\subsection{Synchrotron Field} \label{subsec:synch}
In order to include the synchrotron field, we need an estimate not only for the total synchrotron emission but for the synchrotron energy density seen by a photon at each point down the jet, including angular dependence. At each point in the jet there will be synchrotron emission because there is a magnetic field and a population of electrons, but some parts of the jet will emit much more brightly than others. A photon in the jet frame will see an isotropic distribution of synchrotron photons emitted in its vicinity, as well as an an-isotropic distribution of photons emitted by the rest of the jet. To get an estimate for the synchrotron emission, as well as to test the self-consistency of our overall AGN fields and jet model, we use the \texttt{agnpy} package\footnote{\url{https://doi.org/10.5281/zenodo.4687123}} to calculate a spectral energy distribution (SED; Fig. \ref{fig:sed}). We model the jet with multiple spherical blobs lined up along the jet axis, each with a radius, field-strength, electron density and bulk Lorentz factor taken from our jet model described in Sec. \ref{subsec:jet}. Every blob is given a power law electron spectrum with an exponential cutoff in energy, $N_e(E)=\kappa E^{-\beta}\exp(-E/E_{c})$. Throughout the jet, the electron population is assumed to have a cutoff of $E_c=$ 1.3 MeV (see Table \ref{tab:zones}). The synchrotron emission from these blobs is shown in Fig. \ref{fig:sed} as the red line. \par
Gamma-ray emission from blazars is expected to be quite localised. Within the Potter \& Cotter framework, which is a steady-state model, almost all the inverse-Compton (IC) emission comes from the transition region between the parabolic and conical parts of the jet where the electrons come into rough equipartition with the magnetic field. Electrons here are accelerated, perhaps by a standing shock, though the precise acceleration mechanism is not modelled. Similarly, we accelerate electrons in the blob at $r_{\mathrm{vhe}}$ to a cutoff energy of $E_{cut}=$ 1.3 GeV. We then calculate the synchrotron and (steady-state) inverse-Compton emission from this blob using our photon fields\footnote{\texttt{agnpy} uses a ring torus model instead of our elliptical cross-section model. We have taken the ring radius to be the midpoint of our torus. All other photon field models and parameters are the same.}. The orange and brown lines in Fig. \ref{fig:sed} show this synchrotron and IC emission respectively. The IC emission agrees quite well with the observed steady-state gamma-ray emission. At this large distance ($r_{\mathrm{vhe}}\sim 60$ pc) most of the seed photons for IC are from the torus. \par
The width of the jet at the steady-state emission region is too large to undergo causal variations on flare timescales. Rather than assume a large-scale change in the jet properties, we assume the flare emission is caused by a localised acceleration of electrons in the parabolic base, as might be expected from, e.g., magnetic re-connection or magnetoluminescence (e.g., \cite{Blandford_2017, Uzdensky_2010, Sironi_2014}). We model the flare emission region as a smaller blob ($R_{blob}=R_{jet}/2.8$) in the highly magnetised base of the jet, where the electrons are accelerated up to $E_c=$ 250 MeV. For the location of this flare emission region, we use the lower limit from Ref. \cite{Meyer_Scargle_Blandford_2019}, $r_{\mathrm{em}} = 0.1$ pc. The synchrotron and IC emission from this region are shown as the green and pink lines in Fig. \ref{fig:sed} respectively. The gamma-ray emission, mostly off the BLR at this $r$, is very consistent with the observed flaring spectrum, but the x-ray emission is slightly low.
The overall synchrotron emission -- adding up the emission from all blobs down the jet -- is largely consistent with observations. Therefore, our jet and field models seem to be reasonable.

\begin{figure}[t]
  \centering
    \includegraphics[width=0.48\textwidth]{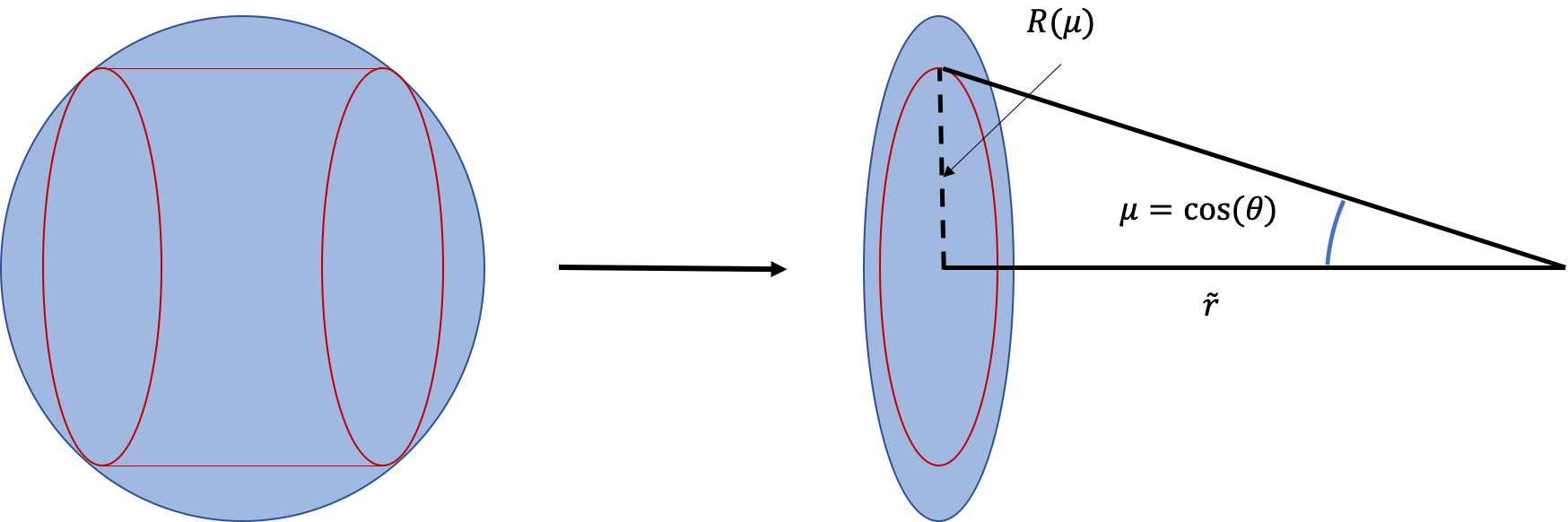}
  \caption{Diagram to show the solid angle approximation used for the flux from the synchrotron blobs. Each blob not containing $r$ is flatten so each $\mu$ corresponds to one radius $R(\mu)$. The energy density is then weighted at each $\mu$ by the area of the red cylinder -- a slice through the original un-flattened blob.}
  \label{fig:flat_blob}
\end{figure}
\subsubsection{In the jet frame} \label{subsubsec:jetframe}
The \texttt{agnpy} model gives the observed synchrotron emission from each blob down the jet (Fig. \ref{fig:sed}),
\begin{equation}\label{eq:vFv_obs}
    \nu F_{\nu}^{obs}(\nu_{obs})=\frac{L(\nu')}{4\pi d_L^2}\delta_s(r)^3,
\end{equation}
at observed frequencies $\nu_{obs}=\delta_s\nu'$, where $\delta_s(r)$ is the Doppler factor of the source ($\theta_s=0.8^{\circ}$), $d_L$ is the luminosity distance to the source, and $L(\nu')$ is the synchrotron luminosity in the blob frame.
For the $\chi$ calculation, we need the total synchrotron photon field in the jet frame at each point along the jet. First, the observed fluxes can be transformed into the blob frame for each blob,
\begin{equation}\label{eq:vFv_blob}
    \nu F_{\nu}^{blob}(\nu')=\frac{L(\nu')}{4\pi R_{blob}^2}=\frac{\nu F_{\nu}^{obs}(\nu_{obs})}{\delta_s^3} \Big(\frac{d_L}{R_{blob}}\Big)^2.
\end{equation}
Then the flux from this blob will be seen at a point $r$ down the jet as,
\begin{equation}\label{eq:vFv_r}
    \nu F_{\nu}^{r}(\nu_r)=\frac{L(\nu')}{4\pi \tilde r^2} \Gamma_{eff}^3 = \nu F_{\nu}^{blob}(\nu') \Big(\frac{R_{blob}}{\tilde r}\Big)^2\Gamma_{eff}^3,
\end{equation}
where $\tilde r = |r-r_{blob}|$, and $\Gamma_{eff}$ is the effective bulk Lorentz factor of the blob when it is observed from $r$, which depends on the difference in bulk jet velocity, $\beta_{eff}=\beta(r) - \beta(r_{blob})$, $\Gamma_{eff}=(1-\beta_{eff}^2)^{-\frac{1}{2}}$, and shifts the frequency, $\nu_r = \nu'\Gamma_{eff}$. Because all of the jet is moving very relativistically ($\Gamma_{bulk}>27$), $\Gamma_{eff}$ is always very close to 1 (within $1\times10^{-7}$). For simplicity we set $\Gamma_{eff}=1$ always, which means $\nu_r = \nu'$.\par
A given flux can then be converted into spectral energy density by dividing by $c$, $u(\nu)=\nu F_{\nu}(\nu)/\nu c$. The total energy density seen at a point $r$ will be the blob-frame energy density of the blob containing $r$ (from Eq. \eqref{eq:vFv_blob}) plus the energy density of all the other blobs as seen from $r$ (from Eq. \eqref{eq:vFv_r}).

Equation \eqref{eq:chi_prime} ($\chi(\omega')$) also requires the solid angle dependence of the energy density. The field from the surrounding blob (indexed by $s$ in what follows) will be isotropic. For speed of computation, we flatten every other blob (indexed by $i$) to a disk (as shown in Fig. \ref{fig:flat_blob}), and assume there is no radial energy dependence of the blob emission. This means for each blob $i\neq s$,
\begin{equation}
    u^i(\nu,\Omega)=u^i(\nu)P^i(\mu)/2\pi.
\end{equation}
Looking at a given blob, each $\mu$ corresponds to a radius $R(\mu)=\tilde r \sqrt{\mu^{-2}-1}$ on the flattened disk. We weight the energy density by the area of a cyclindrical slice of radius $R(\mu)$ through the blob (see Fig. \ref{fig:flat_blob}),
\begin{equation}\label{eq:P_mu}
    P(\mu) = \kappa_0 R(\mu) \sqrt{R_{blob}^2-R(\mu)^2},
\end{equation}
with normalisation,
\begin{equation}\label{eq:kappa}
    \kappa_0 = \Big[\int^{\mu_{max}}_{\mu_{min}}R(\mu) \sqrt{R_{blob}^2-R(\mu)^2} d\mu \Big]^{-1}.
\end{equation}
 As $\Gamma_{eff}=1$, the angles subtended by the blobs do not need to be transformed. Finally, then
\begin{equation}\label{eq:synch_u}
    \begin{aligned}
    u(\nu',\Omega;r)' = \frac{1}{4\pi\nu'c}\Big[\nu F_{\nu}^{blob}(\nu') +
    2 \sum_{i\neq s}\nu F_{\nu}^{i}(\nu') P^i(\mu) \Big],
    \end{aligned}
\end{equation}
where $F_{\nu}^{blob}(\nu')$ is the isotropic emission in the blob containing $r$, and the other blobs each have a position $r_i$ and a radius $R_{blob}^i$. $u(\nu,\Omega)$ can again be converted to $u(\epsilon, \Omega)$ by diving by $h$.

\subsection{Total energy densities}\label{subsec:u_tot}
\par
\begin{figure}
  \centering
    \includegraphics[width=0.48\textwidth]{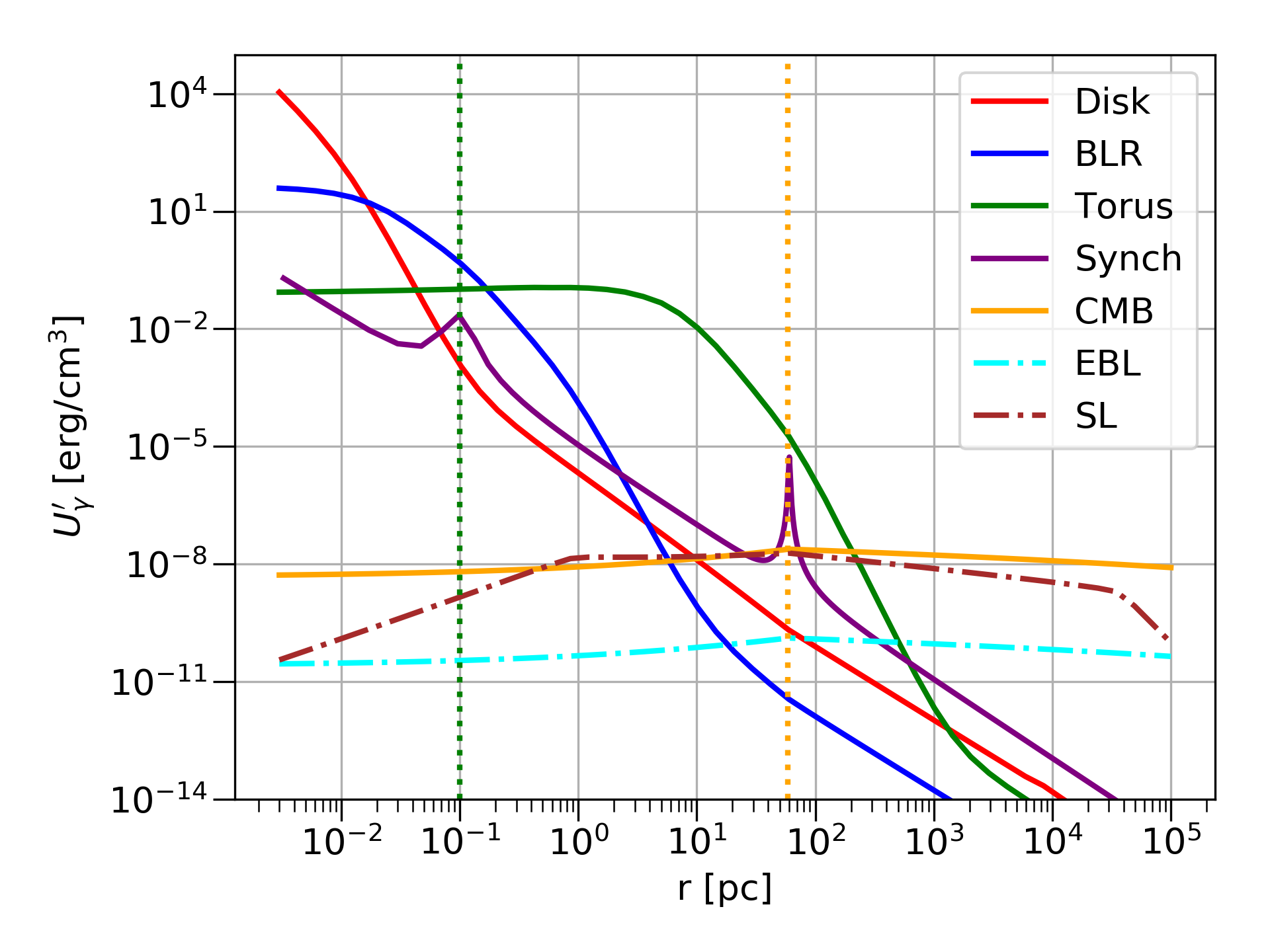}
  \caption{Total photon energy densities (Eq. \eqref{eq:u_tot}) in the co-moving frame of the jet at each $r$. Each field (apart from the EBL or starlight, SL) dominates the total energy density at different radii. The green vertical line shows $r_{\mathrm{em}}$ and the orange vertical line shows $r_{\mathrm{vhe}}$.}
  \label{fig:u_tot}
\end{figure}

 The overall photon field energy densities,

\begin{equation}\label{eq:u_tot}
    U_{\gamma}'(r) = 2\pi \int_{0}^{\infty}d\epsilon'\int_{-1}^{1}d\mu'u'(\epsilon',\Omega';r),
\end{equation}
integrated over energy and solid angle, in the co-moving frame of the jet are shown in Fig. \ref{fig:u_tot}. Each of the AGN fields dominates when $r$ is on the scale of $R_{\mathrm{field}}$, beyond which it becomes strongly de-boosted and begins to look like a point source. Beyond a few 100 pc the boosted CMB field then dominates, just above the starlight. The synchrotron field never quite dominates the overall photon field energy density, but comes close very briefly at the steady-state emission region.

\begin{figure}
  \centering
    \includegraphics[width=0.48\textwidth]{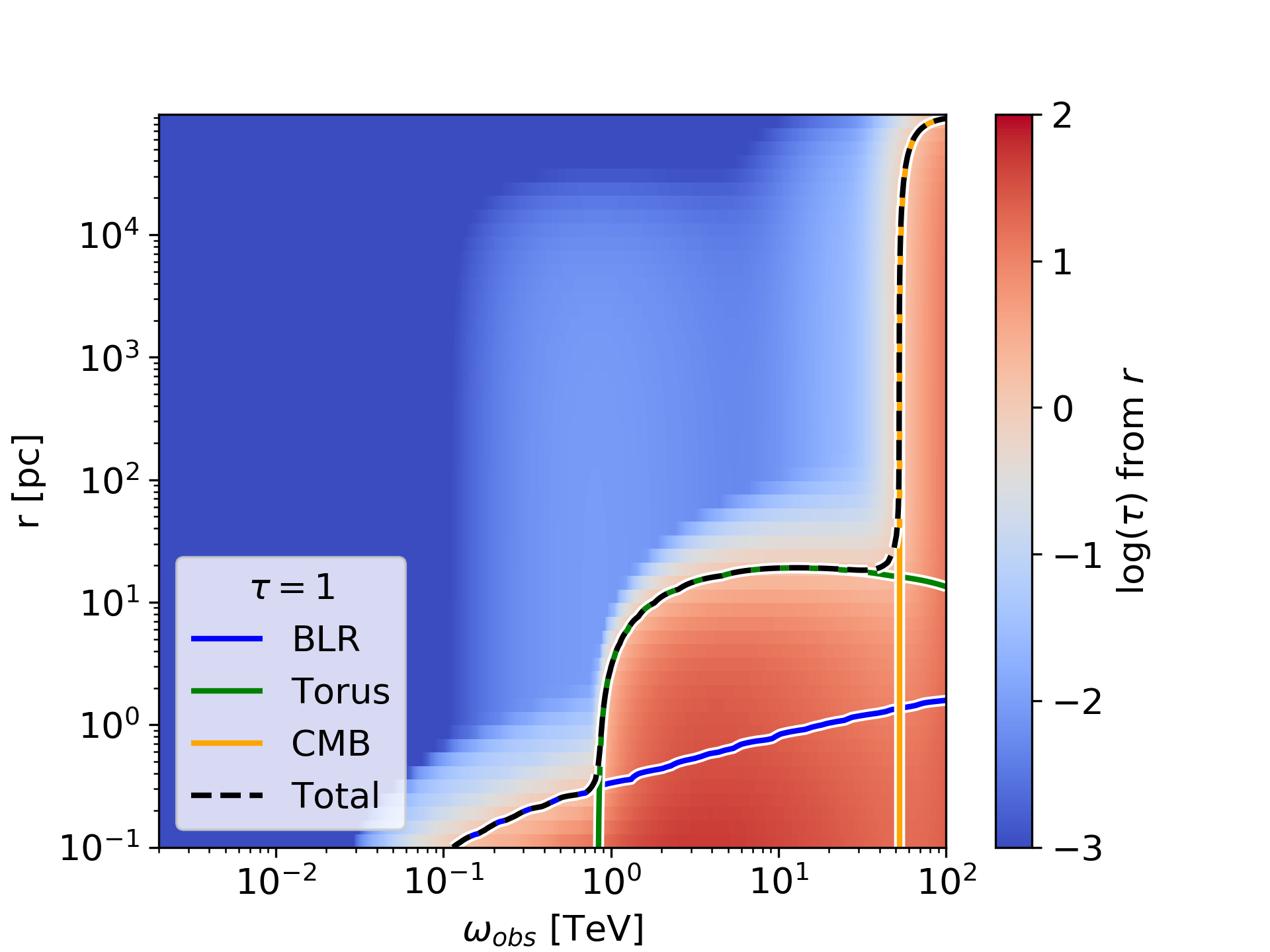}
  \caption{Optical depth, $\tau$ vs $r$ for pair-production with the background photon fields. $\tau$ is calculated by integrating the pair-production absorption rate over the whole jet from $r$, so it shows the optical depth at a given energy for an emission region placed at $r$. Energies are in the observed frame, i.e., taking into account redshift and Lorentz transforms from the ($r$-dependent) comoving jet frame. Lines show the $\tau=1$ contours for absorption from the individual fields that reach $\tau=1$ -- the BLR (blue), torus (green), and CMB (orange) -- and for the total from all the fields (black dashed).}
  \label{fig:tau}
\end{figure}

\section{Results} \label{sec:chi}
\subsection{Absorption}\label{subsec:abs}
We are now in a position to calculate the absorption and dispersion terms for the individual photon fields. Figure \ref{fig:tau} shows the optical depth, $\tau$, seen by a gamma-ray of observed energy $\omega_{obs}$ emitted at $r$, calculated using Eq. \ref{eq:tau_obs}. The coloured contours show where $\tau=1$ for absorption from each of the fields that reach $\tau=1$: the BLR (blue), the torus (green), and the CMB (orange). The black dashed line is the total $\tau=1$ contour, including all the background fields.\par
\begin{figure}
  \centering
    \includegraphics[width=0.48\textwidth]{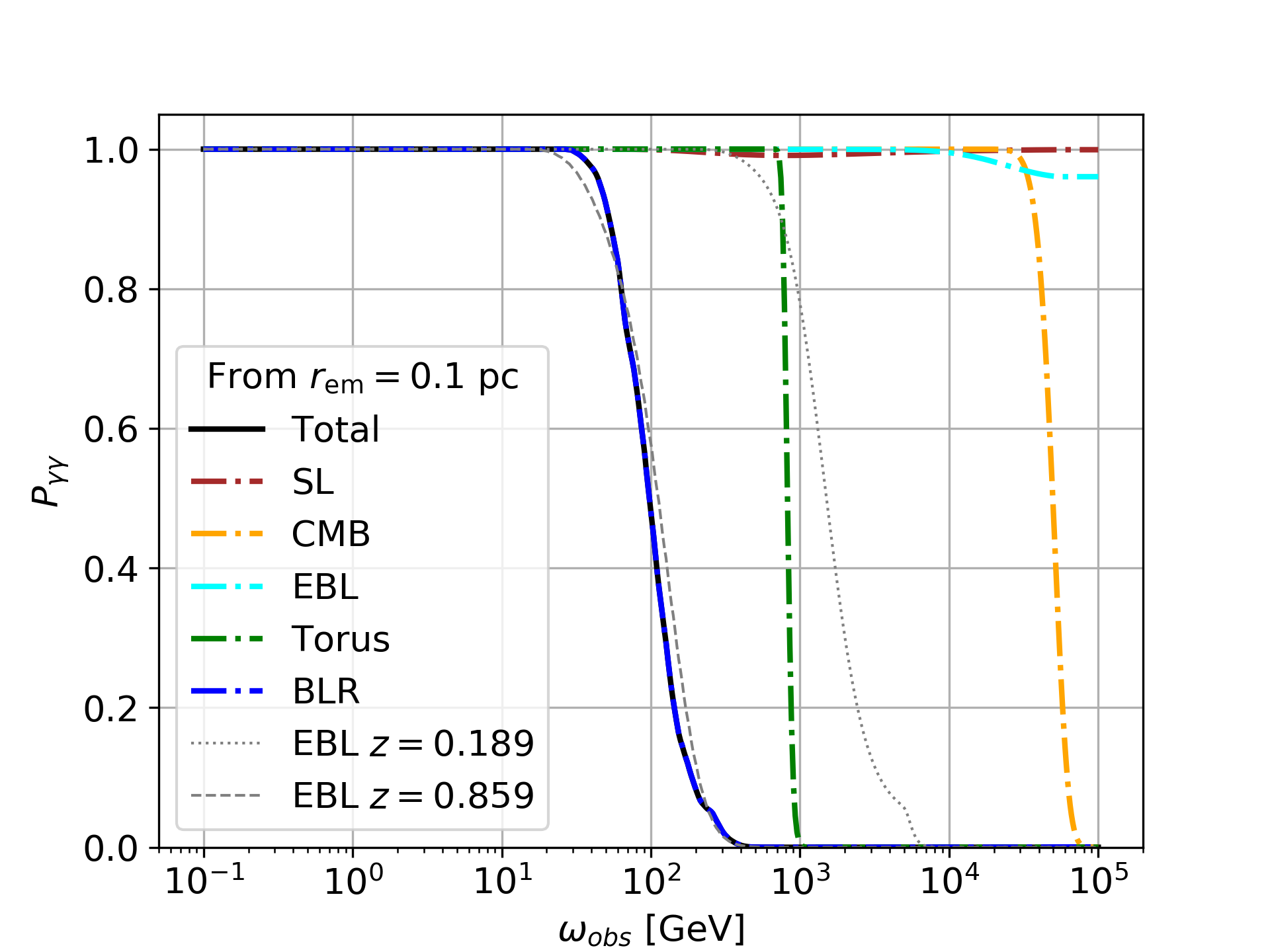}
    \includegraphics[width=0.48\textwidth]{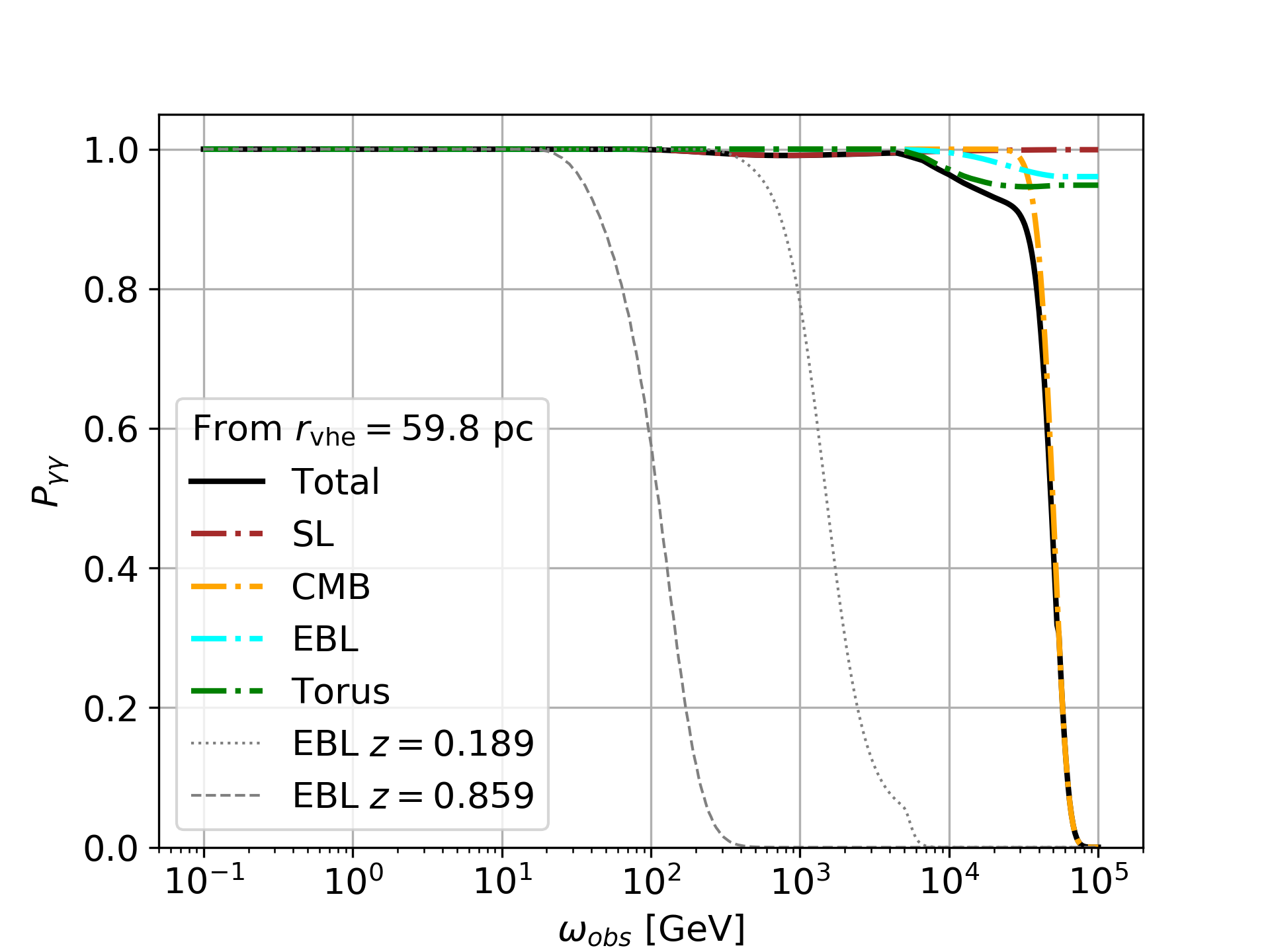}
  \caption{Photon survival probabilities, $P_{\gamma\gamma}$, vs. observed energy only including absorption from the background fields within the jet, for our two emission regions, $r_{\mathrm{em}}$ (top) and $r_{\mathrm{vhe}}$ (bottom). Coloured lines show the absorption from the individual fields and the black lines show the total. Dashed grey lines show the EBL absorption in intergalactic space for two redshifts for comparison, $z=0.859$ (3C454.3) and $0.189$ (PKS 0736+017, the closest FSRQ).}
  \label{fig:abs_pggs}
\end{figure}

Figure \ref{fig:abs_pggs} shows the photon survival probabilities, $P_{\gamma\gamma}$, vs. observed energy for gamma rays propagated from our two emission regions--$r_{\mathrm{em}}$ (top) and $r_{\mathrm{vhe}}$ (bottom)--only including absorption within the jet (so $P_{\gamma\gamma}=\exp(-\tau)$ in this case). Again, the coloured lines show the absorption from the individual fields, and the black line shows the total. Photons from the BLR, torus, and CMB each dominate gamma-ray absorption at progressively larger distances down the jet and at progressively higher energies. Absorption by starlight is always negligible. For gamma rays emitted from $r_{\mathrm{em}}=0.1$ pc, absorption from BLR photons dominates.  This absorption starts at around $45$ GeV (around the energy of intergalactic EBL absorption for 3C454.3), whereas torus absorption would only begin at around $800$ GeV, and CMB absorption at around $30$ TeV. Gamma rays emitted from the steady state emission region $r_{\mathrm{vhe}}=59.8$ pc, however, can avoid any significant absorption up to CMB absorption energies. This threshold would increase for sources at lower redshift (where the CMB energy density is lower) but it is already a long way above the energy of intergalactic EBL absorption (shown by the dashed grey lines in Fig \ref{fig:abs_pggs}), even for PKS 0736+017, the closest FSRQ.\par
 These energies define the energy ranges where it is sensible to discuss the effects of dispersion.

\begin{figure}
  \centering
    \includegraphics[width=0.48\textwidth]{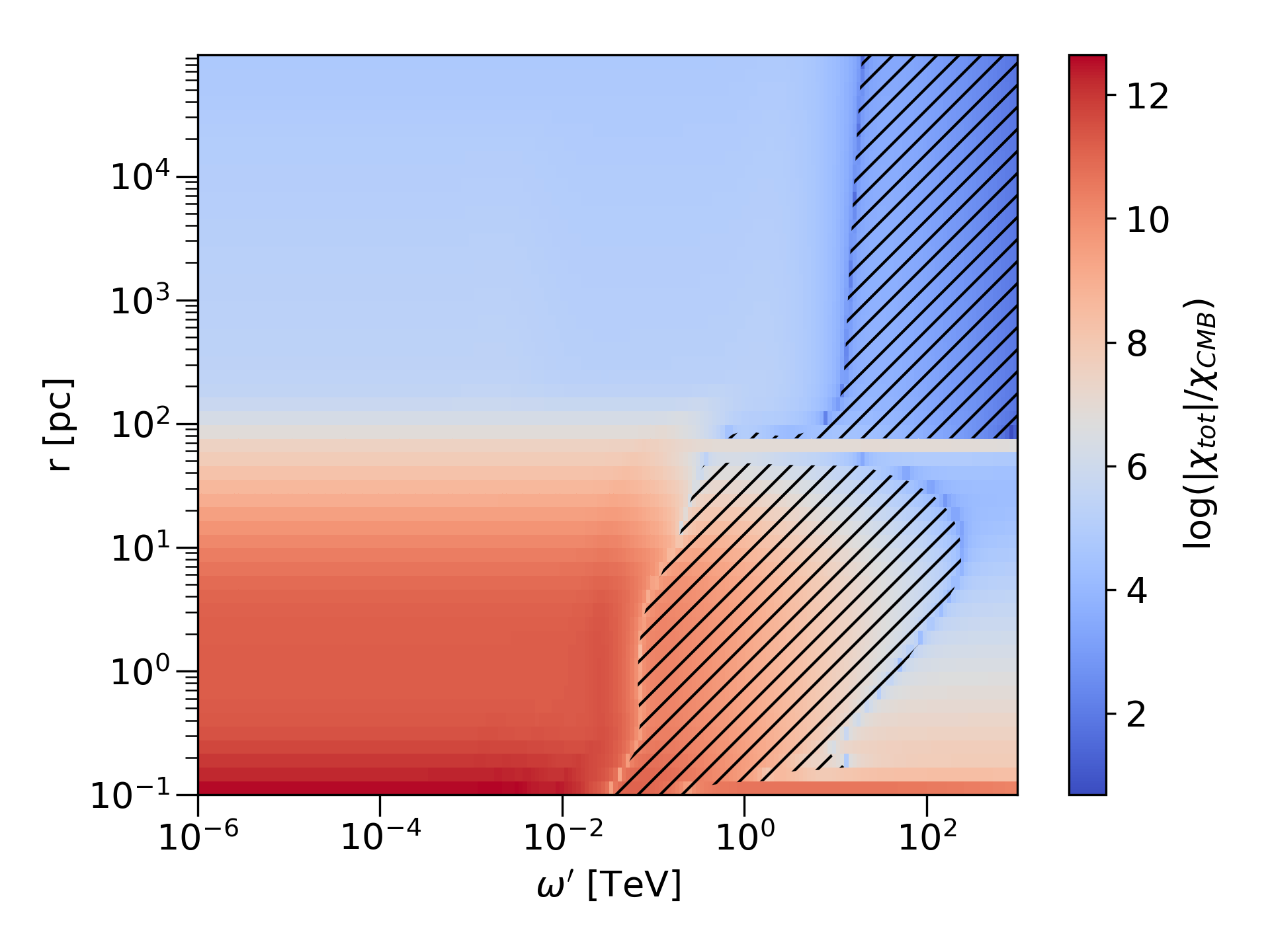}
  \caption{Total $\log(|\chi_{tot}|/\chi_{CMB})$ vs $\omega'$ (the photon energy in the jet frame) and $r$. Absolute values are shown, the hatched region corresponds to negative $\chi$.}
  \label{fig:chi_tot}
\end{figure}

\subsection{Dispersion}\label{subsec:chi}
Figure \ref{fig:chi_tot} shows $\chi_\mathrm{tot}(\omega')/\chi_{CMB}$ evaluated at  distances $r$ down the jet (by performing the integral in Eqn. \ref{eq:chi_prime}, including all photon fields). $\chi_{CMB}=5.11\times10^{-43}$ is the low energy (far from pair production energies) value of $\chi$ for the isotropic CMB field at $z=0$ (see Sec. \ref{sec:fields}). For the whole jet, $\chi$ is greater than $\chi_{CMB}$ at most energies by at least a factor of 100, and up to a factor of about 10$^{12}$. As in Ref. \cite{Dobrynina_2015}, $\chi(\omega)$ for a given field is constant at low energies, peaks around the pair production energy, and then rapidly becomes negative before returning to zero. The energy $\omega$ at which the pair production threshold is reached is different for the different fields (see Sec. \ref{subsec:abs} above). $\chi(\omega)$ will therefore peak at different energies for each field. If $\omega$ has just exceeded the pair production threshold for the dominating field, the total, $\chi_\mathrm{tot}$, can be negative.\par
\begin{figure}
  \centering
    \includegraphics[width=0.48\textwidth]{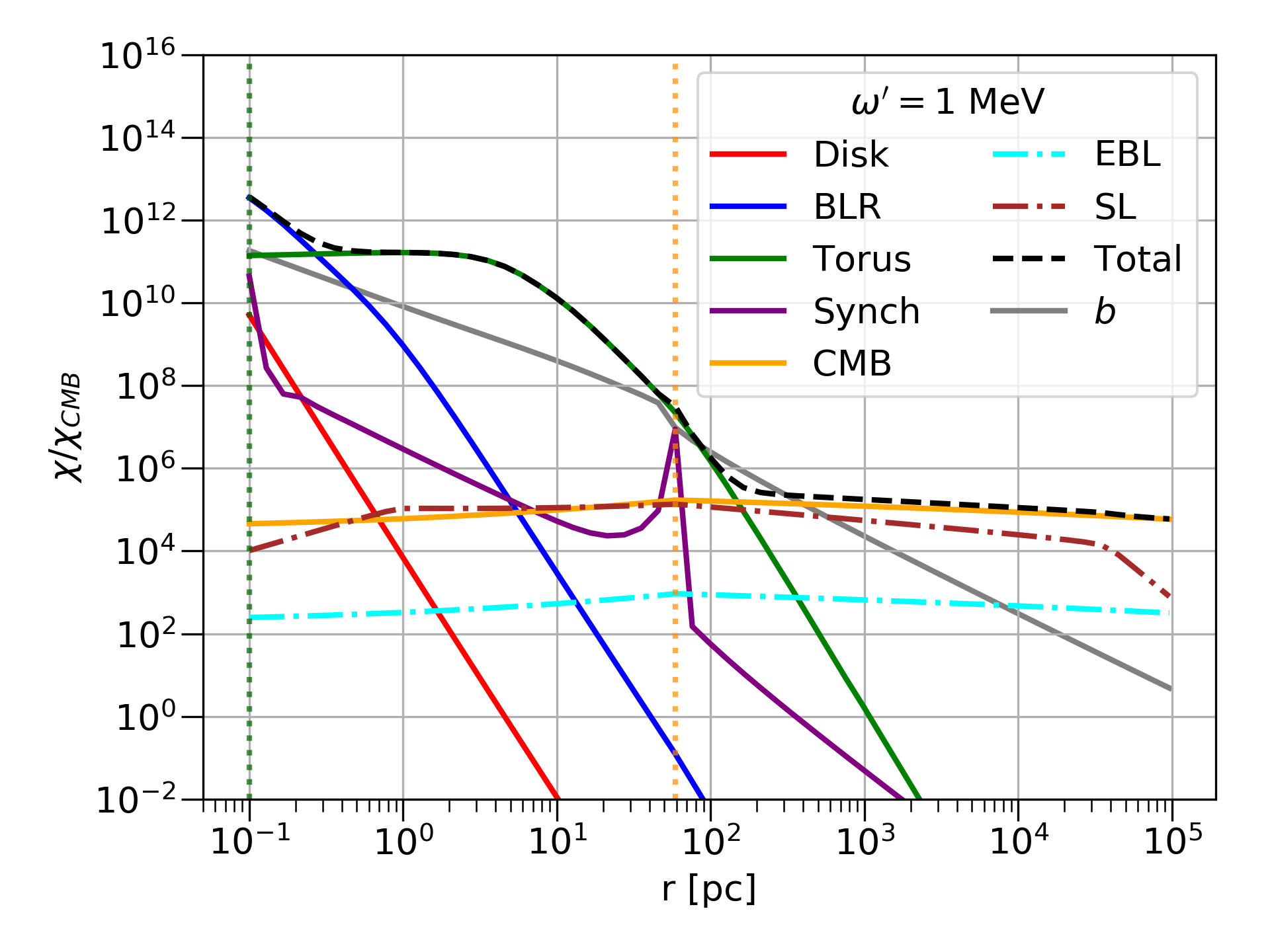}
  \caption{Total $\chi$ (as a ratio of $\chi_{CMB}$) vs $r$ in the jet frame for a low energy photon of $1$ MeV. The BLR, torus and boosted CMB fields dominate for increasing $r$. The synchrotron field is important around $r_{\mathrm{vhe}}$. Dispersion off the magnetic field (defined in Eq. \eqref{eqn:b}) is shown in grey for comparison. The green vertical line shows $r_{\mathrm{em}}$ and the orange vertical line shows $r_{\mathrm{vhe}}$.}
  \label{fig:chi_r}
\end{figure}

Figure~\ref{fig:chi_r} shows $\chi$ vs $r$ for $\omega'=$ 1 MeV, with a breakdown of the contributions from the different fields. The main fields that contribute are the BLR, torus, and CMB fields which dominate at progressively larger radii respectively, where they each dominate the overall energy density (Fig.  \ref{fig:u_tot}) and absorption (Fig. \ref{fig:tau}). Dispersion off the CMB is not constant and is not equal to $\chi_{CMB}$ everywhere because in the jet frame it is boosted and is not at $z=0$. The synchrotron field briefly comes close to dominating at $r_{\mathrm{vhe}}$ at all energies, and dominates at small $r$ only at very high energy $\omega'\gtrsim100$ TeV  (the bottom right of figure \ref{fig:chi_tot}) because the pair production threshold off of the synchrotron photons is so high.
\subsection{Effect on photon survival probability}\label{subsec:pggs}
\begin{figure}
  \centering
    \includegraphics[width=0.48\textwidth]{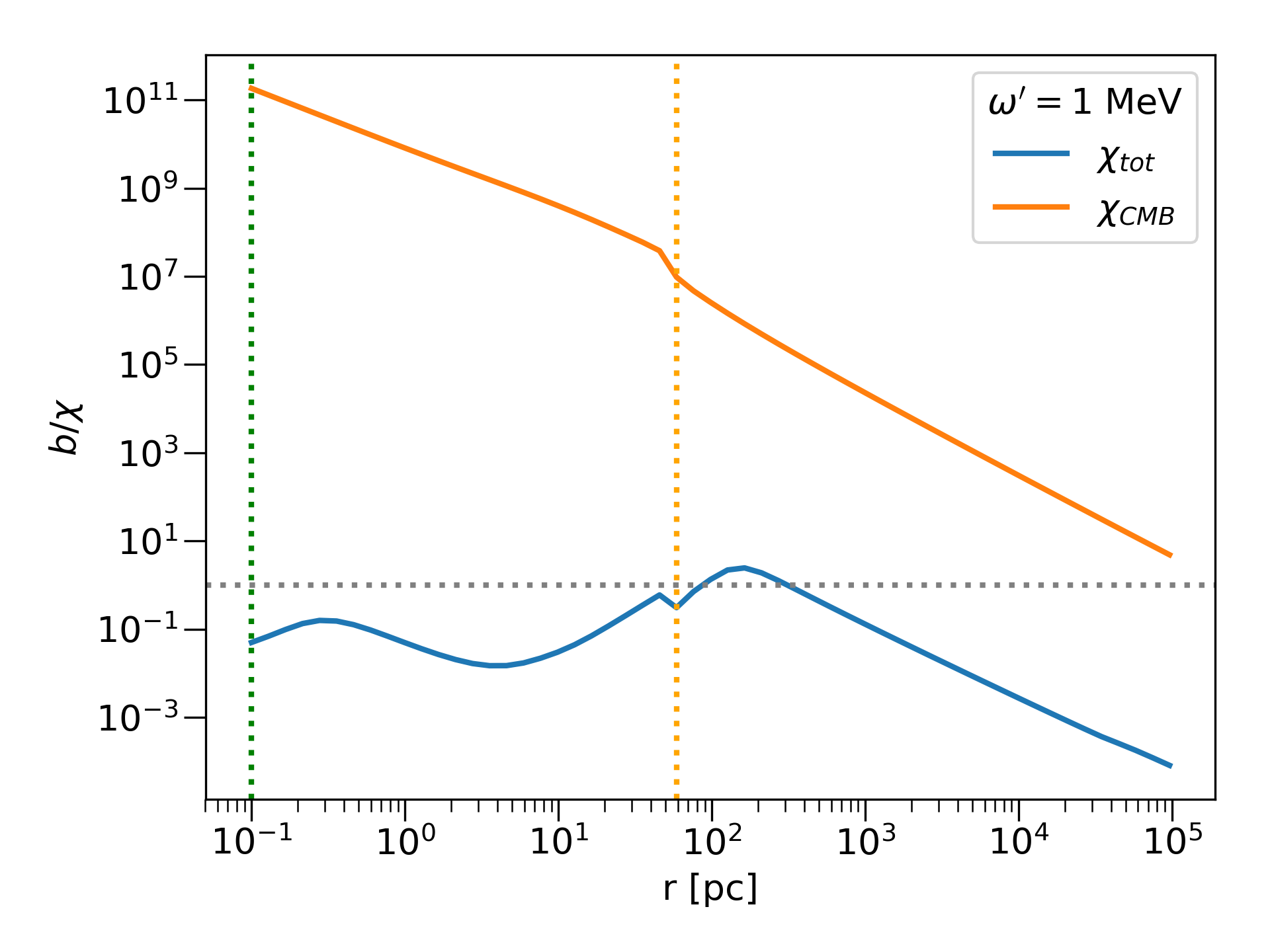}
  \caption{Ratio $b/\chi$ as a function of $r$ for a low energy photon of 1 MeV. The green vertical line shows $r_{\mathrm{em}}$ and the orange vertical line shows $r_{\mathrm{vhe}}$. $\chi$ becomes important when $\chi>b$ (below the horizontal dashed line).}
  \label{fig:b_over_chi}
\end{figure}

For a photon of energy $E$ propagating in a homogeneous, transverse field, $B$, the wave-number for ALP-photon oscillations into an ALP with mass $m_a$ and coupling $g_{a\gamma}$ is given by,
\begin{equation}\label{eqn:delta_osc}
    \Delta_{osc} = \sqrt{\left\{\frac{|m_a^2 - m_T^2|}{2E} + E\left(b + \chi + i \frac{\Gamma_{\gamma\gamma}}{2}\right)\right\}^2 + (g_{a\gamma}B)^2},
\end{equation}
where $m_T$ is the effective mass of the photon (see Ref. \cite{Davies_2021}), $\chi$ and $\Gamma_{\gamma\gamma}$ are the total dispersion and absorption terms for the surrounding photon fields respectively, and,
\begin{equation}\label{eqn:b}
    b=\frac{\alpha}{45\pi}\Big(\frac{B}{B_\mathrm{cr}}\Big)^2
\end{equation}
is the vacuum QED term describing dispersion off of the magnetic field, with $B_\mathrm{cr}$ the critical magnetic field $B_{cr}=m_e^2/|e|\sim4.4\times10^{13}$ G, where $e$ is the electric charge. To focus on the effects of dispersion, in the following discussion we assume that $E$ is such that $\tau(E)<<1$ for all photon fields, so we can ignore the $\Gamma$ term. Eq. \eqref{eqn:delta_osc} then leads to two `critical energies', around which the ALP-photon oscillation length depends strongly on energy, leading to oscillations in energy spectra:
\begin{equation}\label{eq:e_low}
    E_\mathrm{crit}^\mathrm{low} = \frac{|m_a^2 - m_T^2|}{2g_{a\gamma}B}
\end{equation}
which depends only on the effective mass difference between the ALP and the photon and is independent of $\chi$, and
\begin{equation}\label{eq:e_high}
    E_\mathrm{crit}^\mathrm{high} = \frac{g_{a\gamma}B}{b+\chi}.
\end{equation}
Between these two energies (if $E_\mathrm{crit}^\mathrm{low}<E_\mathrm{crit}^\mathrm{high}$) is the so-called `strong mixing regime', where photons are maximally converted into ALPs with an oscillation length independent of energy. $\chi$ can affect ALP-photon mixing by affecting $E_\mathrm{crit}^\mathrm{high}$. Because a magnetic field is required for any mixing at all, $b$ is always greater than 0. Therefore, whether $\chi$ will have an effect or not depends on the comparative sizes of $b$ and $\chi$. If $b\gg\chi$, $\chi$ will not affect mixing and $E_\mathrm{crit}^\mathrm{high}$ will remain independent of changes in $\chi$. On the other hand, if $\chi\gg b$, $E_\mathrm{crit}^\mathrm{high}$ will become $\propto 1/\chi$ and the value of $\chi$ will be important.
Figure \ref{fig:b_over_chi} shows $b/\chi$ for $\chi_{CMB}=5.11\times10^{-43}$ and for $\chi_{tot}$ down the jet, from $r_{\mathrm{em}}$, for a low energy $\omega'=1$\,MeV, when all the fields are taken into account. Below the dotted line, $\chi>b$ and so $\chi$ starts to become important. $\chi_{tot}$ is comparable to or larger than $b$ until about $r_{\mathrm{vhe}}$ and then becomes much larger than $b$ as $B$ decreases down the jet, whereas $\chi_{CMB}$ only becomes comparable to $b$ in the very weak field at the end of the jet. This implies that the full photon-photon dispersion calculation could be important within the jet.

\begin{figure*}
  \centering
    \includegraphics[width=\textwidth]{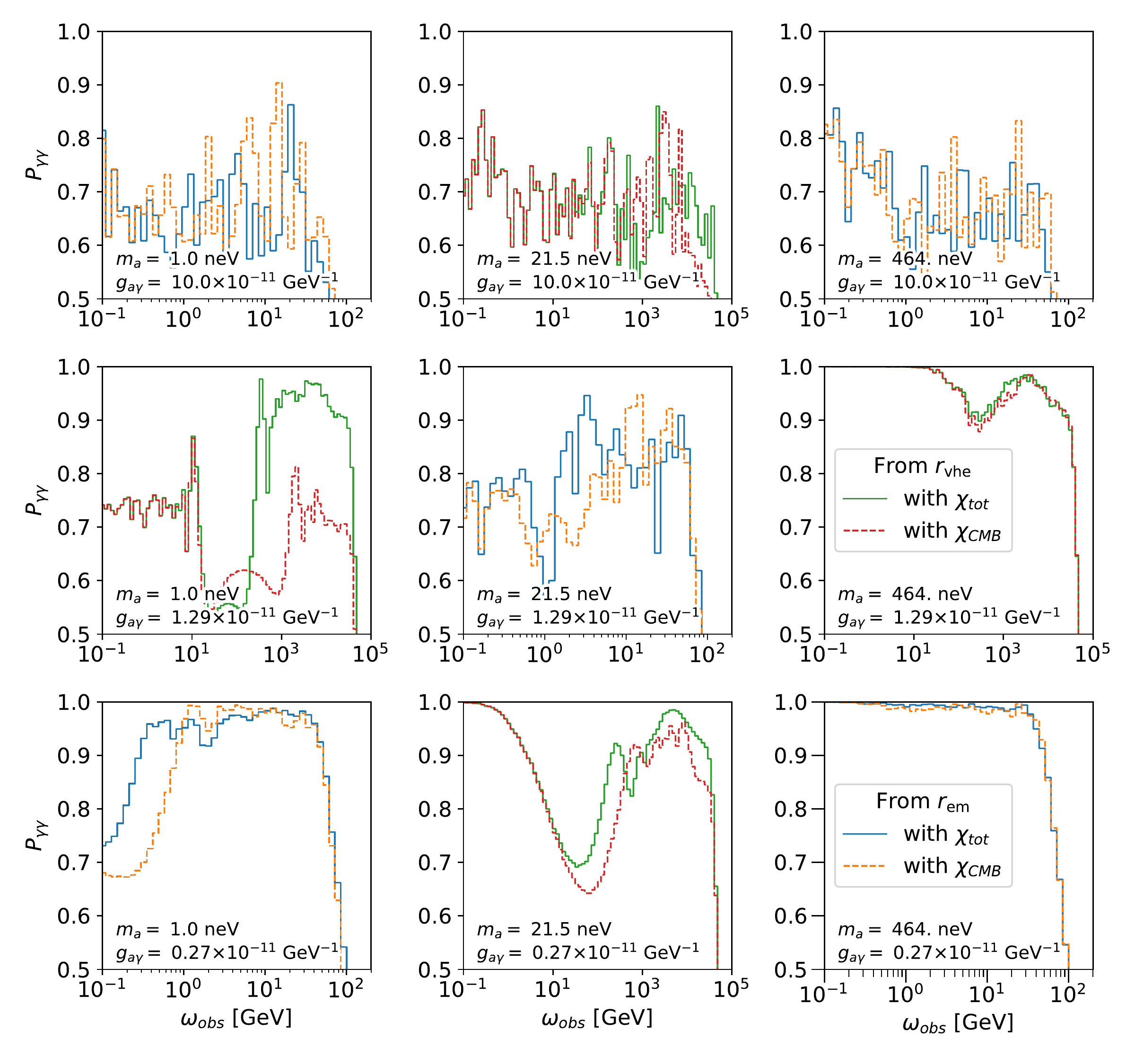}
  \caption{Photon survival probabilities, $P_{\gamma\gamma}$, for different values of ALP mass, $m_a$, and coupling, $g_{a\gamma}$, with and without $\chi_{tot}$. The plots are arranged with increasing $m_a$ from left to right, and decreasing $g_{a\gamma}$ from top to bottom. ALP-photon beams are propagated through one realisation of our jet, from either $r_{\mathrm{em}}$ or $r_{\mathrm{vhe}}$: Blue and orange-dashed lines show propagation from $r_{\mathrm{em}}$ with and without $\chi_{tot}$ respectively; Green and red-dashed lines show propagation from $r_{\mathrm{vhe}}$ with and without $\chi_{tot}$ respectively. $30 \%$ of the field is in a tangled component (as described in Ref. \cite{Davies_2021}). The $P_{\gamma\gamma}$s are averaged over roughly Fermi-sized energy bins (16 per decade), though the energy range extends beyond Fermi energies to 100 TeV for the $r_{\mathrm{vhe}}$ plots.}
  \label{fig:pggs}
\end{figure*}

\begin{figure}
  \centering
    \includegraphics[width=0.48\textwidth]{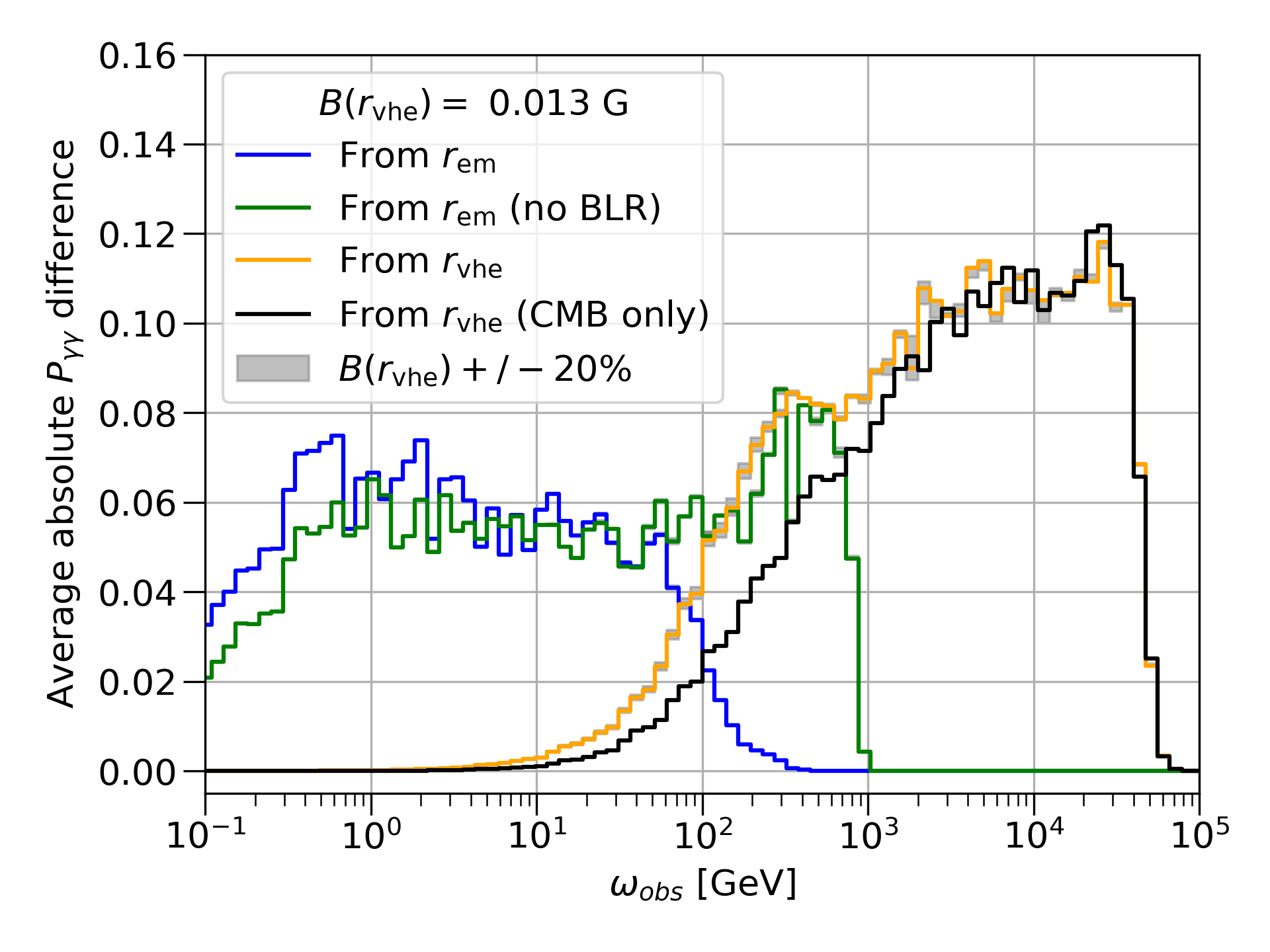}
  \caption{Average absolute differences in $P_{\gamma\gamma}$ between using $\chi_{tot}$ and just $\chi_{CMB}$ vs observed energy for 100 logarithmically spaced points in ($m_a$, $g_{a\gamma}$)-space between $m_a=1 - 1000$ neV and $g_{a\gamma}=0.1 - 10 \times10^{-11}$ GeV$^{-1}$. The blue line is for ALPs propagated from $r_{\mathrm{em}}$, including all photon fields. The green line is from $r_{\mathrm{em}}$ but not including the BLR. The orange line is from $r_{\mathrm{vhe}}$, including all fields (steady-state), and the black line is with the CMB only. Dispersion has the largest effect above 100 GeV. The difference when $B$ adjusted by 20\% also shown as grey shaded regions for each line.}
  \label{fig:av_diff}
\end{figure}

Figure \ref{fig:pggs} shows some photon survival probabilities, $P_{\gamma\gamma}$, for various values of ($m_a$, $g_{a\gamma}$) with and without $\chi_{tot}$. The ALP-photon beams are propagated through one realisation of our jet, with $30 \%$ of the field in a tangled component (as described in Ref. \cite{Davies_2021}). In both cases we include absorption from the photon fields in the ALP-photon mixing equations, hence the exponential cutoffs. Plots with blue and orange-dashed lines are for propagation from $r_{\mathrm{em}}$ and plots with green and red-dashed lines are for propagation from $r_{\mathrm{vhe}}$. The two cases therefore show absorption at different energies (cf. Fig. \ref{fig:abs_pggs}). The $P_{\gamma\gamma}$s are averaged over energy bins which are roughly Fermi-sized (16 bins per decade), though the energy range extends beyond Fermi energies to 100 TeV for the $r_{\mathrm{vhe}}$ plots. As can be seen from the figure, $\chi$ can have a large effect on $P_{\gamma\gamma}$. \par
Figure \ref{fig:av_diff} shows the average (over $m_a$ and $g_{a\gamma}$) absolute difference in $P_{\gamma\gamma}$ with and without $\chi_{tot}$ for a $10\times10$ logarithmically spaced grid in ($m_a$, $g_{a\gamma}$)-space between $m_a=1 - 1000$ neV and $g_{a\gamma}=0.1 - 10 \times10^{-11}$ GeV$^{-1}$. The different lines show the differences when ALPs are propagated from the two emission regions, including various fields. Again, absorption is included in both cases (with and without $\chi_{tot}$) and therefore the differences go to zero above the relevant absorption energies. In general, the differences are largest at the highest energies. This is because the effect of $\chi$ is to lower $E_\mathrm{crit}^\mathrm{high}$, often meaning that mixing is strongly reduced at the highest energies as they are now above the maximum energy at which efficient mixing occurs. This could have implications for future searches at very high energies (with, e.g., the Cherenkov Telescope Array \cite{Science_with_CTA_2018, CTA_Gpropa_2021}), potentially reducing the effectiveness of those using blazars as their source, especially those looking for a reduced opacity of the universe at the highest energies.\par
In detail, the blue line in Figure \ref{fig:av_diff} shows the differences for ALPs propagated from $r_{\mathrm{em}}$ (the flare emission region), including all the photon fields. Including the full dispersion calculation causes $\sim6\%$ average differences for all energies up to where BLR absorption becomes important ($\tau=1$ at around $100$ GeV for gamma-rays emitted from $r_{\mathrm{em}}$, see Fig. \ref{fig:tau}). The green line is also for ALPs propagated from $r_{\mathrm{em}}$, but excluding the BLR field--allowing the differences to extend out to $\sim$ TeV, where torus absorption kicks in. This approximates an FSRQ seen at energies above BLR absorption energies, where the emission region must be further out from the BLR. In this case, the differences are not quite as large at low energies (where BLR dispersion dominates), but are still substantial because of dispersion from the torus photons.\par
The orange line in Fig. \ref{fig:av_diff} shows the differences for propagation from $r_{\mathrm{vhe}}$ (the steady state emission region), including all photon fields. In this case, because the very largest $\chi$s are produced by the AGN fields at smaller radii (see Fig. \ref{fig:chi_r}), dispersion does not make a significant difference until around $100$ GeV. Absorption, however, is not important until CMB absorption energies at around $40$ TeV (see Fig. \ref{fig:abs_pggs}). Comparing the green and orange lines shows that dispersion from torus photons is still important briefly until around $1$ TeV. Beyond that energy, it is the boosted CMB field which dominates dispersion -- Fig. \ref{fig:chi_r} shows that the CMB dominates $\chi_{tot}$ for a large amount of the jet ($r\gtrsim100$ pc), and comparison with Fig. \ref{fig:b_over_chi} shows that this is where $b/\chi$ is smallest. Indeed, the black line shows the differences when only the boosted CMB field is included, and it is very similar to the full steady-state case (orange line) above torus-dispersion energies. This also shows that, in this case, the starlight does not contribute strongly and so the detailed modelling of the starlight field is not important -- though for a source at lower redshift the CMB energy density would be lower and so the starlight contribution would likely be greater (cf. Figs. \ref{fig:u_tot} and \ref{fig:chi_r}).\par
The $r_{\mathrm{vhe}}$ and CMB-only cases also roughly approximate a BL Lac-type source, where the AGN photon fields are not present.
Figure \ref{fig:av_diff} also shows (grey shaded regions), for all cases except the CMB-only, the effects of changing the jet magnetic field by $\pm 20\%$, and therefore changing the synchrotron photon field. Because the synchrotron field does not dominate $\chi$, varying the jet magnetic field does not affect $\chi$ very strongly in any case. This means that the modelling of the synchrotron photon field does not need to be extremely precise for either FSRQs or BL Lacs. \par
In all cases then, dispersion off the boosted CMB should be included for observed energies above 100 GeV. And for emission regions at distances comparable to the AGN field scales, the full dispersion calculation should be included at all observable gamma-ray energies.

\section{Conclusions} \label{sec:conclusions}
Photon-photon dispersion off the CMB is known to be an important effect for ALP-photon mixing calculations, but no other photon fields are usually included. Firstly, we have pointed out that the CMB dispersion term should have a $(1+z)^4$ redshift dependence instead of being constant as is commonly assumed. Further, we have investigated the relevance of dispersion off of the background photon fields likely to be found in blazar jets -- the accretion disk, BLR, dust torus, starlight, and synchrotron fields -- motivated by the fact that blazars are common sources for gamma-ray ALP searches and that the jet itself looks like a promising mixing environment for future searches. We have used the bright FSRQ 3C454.3 as an example, modelling the jet within the PC framework---other possibilities for the bulk Lorentz factor and magnetic field profiles within the jet exist (see, e.g., \cite{Ghisellini_Tavecchio_2009}), though we would not expect these to drastically affect our general results. Performing the full $\chi$ calculation within the jet requires an energy- and angular-dependent photon energy density in the comoving jet frame at each point $r$ (which includes doppler boosting the redshift-dependent CMB field). For the starlight, we have modelled the emission as that of a typical large elliptical galaxy, with a spectrum dominated by K-type stars. We have modelled the sources of the AGN fields as a flat accretion disk, a BLR of concentric rings, and a torus with an elliptical cross-section, each emitting isotropically and monochromatically at a given radius. By calculating full SEDs (flaring and steady-state) of 3C454.3 and checking them against observations, we obtained an estimate for the total synchrotron field at each point $r$ -- as well as confirming the self-consistency of our overall jet and field models.
Then, by propagating ALP-photon beams through our jet with and without dispersion off of all these fields (and including absorption from them), we showed that the full dispersion calculation can have a large effect on the $P_{\gamma\gamma}$s. $\chi$ affects ALP-photon mixing by adjusting $E_\mathrm{crit}^\mathrm{high}$, the upper critical energy around which strong mixing occurs. The relative importance of $\chi$ can be found by comparing it to $b$, the equivalent dispersion term due to the magnetic field. When $\chi>b$, the value of $\chi$ becomes significant. Including all the fields, $\chi_{tot}\gtrsim b$ all along the jet, with the BLR field, the torus field, and the boosted CMB dominating at increasing distances. The synchrotron field only contributes strongly to $\chi_{tot}$ at very high energies (pair production off the low-energy synchrotron field occurs at higher energies than the AGN fields), and around the steady-state emission region. The effects of $\chi$ are particularly strong at observed energies above 100 GeV as increasing $\chi$ reduces $E_\mathrm{crit}^\mathrm{high}$. This could have consequences for future ALP searches, particularly those looking for a reduced opacity of the universe at the highest energies (see Sec. \ref{sec:intro}). The effect at high energies is predominantly caused by dispersion off the boosted CMB, and so remains important even in the BL Lac-type case or in the steady state emission case (where the emission region is beyond most of the AGN field scales). This means that the transformed CMB should always be included in the dispersion calculations for observed energies above 100 GeV. Also, while individual photon fields will vary from source to source, for jet models with emission regions on AGN field scales, the full dispersion calculation should be included for all observable gamma-ray energies, up to where absorption from these same fields becomes significant.

\section*{Acknowledgements} \label{sec:acknowledgements}
J.~D. acknowledges an STFC PhD studentship. M.~M.  acknowledges  support from the European Research Council (ERC) under the European Union’s Horizon 2020 research and innovation program Grant agreement No. 948689 (AxionDM) and from the Deutsche Forschungsgemeinschaft (DFG, German Research Foundation) under Germany’s Excellence Strategy – EXC 2121 „Quantum Universe“ – 390833306. G.~C. acknowledges support form STFC grants ST/V006355/1, ST/V001477/1 and ST/S002952/1 and from Exeter College, Oxford.


\appendix
\section{Torus field model}
\label{app:app}

The torus in Ref. \cite{Finke_2016} is modelled as a flat, extended disk, where every radius ($R$) emits at the same energy, $\epsilon_{dt}=2.7\Theta$ which depends on the temperature of the torus, $\Theta$. Allowing the cloud number density within the torus to drop as $R^{-\zeta}$, the differential energy density for this torus model is (where $x$ and $\mu$ are defined as in Sec. \ref{subsec:agn}):

\begin{equation}\label{eq:u_tor_finke}
\begin{split}
    u(\epsilon, \Omega; r) =& \frac{\xi_{dt}L_{disk}}{(4\pi)^2 c R_{eff}}\delta(\epsilon - 2.7\Theta) \\
    & \times \int^{R_2}_{R_1}\frac{dR}{x^2}\Big(\frac{R}{R_1}\Big)^{-\zeta}\delta(\mu - r/x),
\end{split}
\end{equation}
where $\xi_{dt}$ is the fraction of the disk luminosity re-emitted by the disk, $L_{dt}=\xi_{dt}L_{disk}$, $R_1$ and $R_2$ are the inner and outer radii of the disk respectively, and
\begin{equation}\label{eq:tor_Reff}
    R_{eff} =
    \begin{dcases}
    R_1\ln(\frac{R_2}{R_1}),& \text{if } \zeta=1\\
    \frac{R_1 - R_2\Big(\frac{R_2}{R_1}\Big)^{-\zeta}}{\zeta - 1},              &\text{if } \zeta \neq 1.
\end{dcases}
\end{equation}
\par However, unlike the disk and BLR (see Sec. \ref{subsubsec:disk} and \ref{subsubsec:blr}), actual dusty tori are not thought to be thin (e.g., Ref. \cite{Calderone_2012}). As the calculation of $\chi$ depends on the geometry of the fields and not just their energy density (Eq. \eqref{eq:chi}), the height of the torus could possibly affect the dispersion off the torus field quite a lot, especially on scales similar to the height of the torus. Observations indicate that a covering fraction (the fraction of the sky obscured by the torus, from the point of view of the black hole) of around $f_c\sim0.6$ is typical \cite{Calderone_2012}, meaning that the torus height is not insignificant compared to the radius of $\sim$pc to $10$s pc scales.\par
Therefore, we extend the model of Eq. \eqref{eq:u_tor_finke} to include an elliptical cross-section for the torus. The covering fraction of the torus can then be set by varying the ratio of the semi-major ($a$) and semi-minor ($b$) axes of the ellipse.

\begin{figure}
  \centering
    \includegraphics[width=0.48\textwidth]{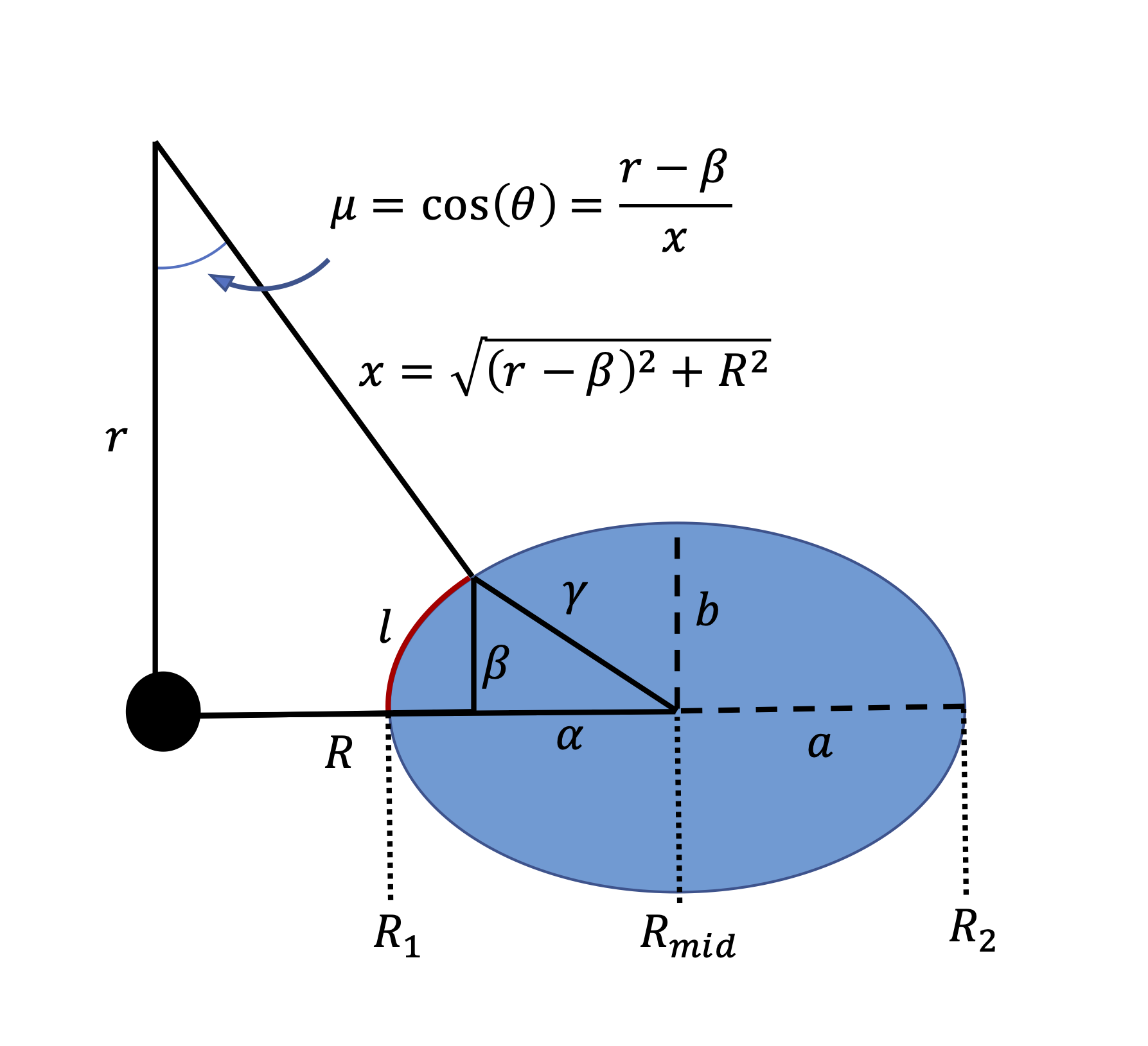}
  \caption{Diagram showing the geometry of the torus model. The torus has an elliptical cross-section. The distance along the jet from the black hole is measure by $r$. As in the disk and BLR models, $R$ measures the radial distance in the plane perpendicular to the jet, and $x$ measures the distance between point $r$ and the background photon emission region--in this case the surface of the torus above point $R$.}
  \label{fig:tor_model}
\end{figure}

Figure \ref{fig:tor_model} shows the geometry of our torus model with an elliptical cross-section. As before, $R$ denotes the radial distance from the black hole in the plane perpendicular to the jet.
We model the emission as coming from the torus surface. The surface of the cross-sectional ellipse is defined by the equation,

\begin{equation}\label{eq:ell}
    \frac{\alpha^2}{a^2} + \frac{\beta^2}{b^2} = 1,
\end{equation}
where $\alpha$ is the distance from the ellipse centre along the semi-major axis, and $\beta$ is the distance from the ellipse centre along the semi-minor axis (see Fig. \ref{fig:tor_model}). A point on the surface is therefore a distance, $\gamma^2 = \alpha^2 + \beta^2$ from the ellipse centre.\par
Viewed from a distance, $r$, above the black hole, a point on the torus surface will be a distance $x$ away, with,
\begin{equation}\label{eq:ell_x}
    x = \sqrt{(r-\beta)^2 + R^2},
\end{equation}
where $R=R_{mid} - \alpha$, and will be seen at an angle,
\begin{equation}\label{eq:ell_mu}
    \mu = \frac{r-\beta}{x},
\end{equation}
as opposed to $\mu=r/x$ for a flat disk.\par
For the integral, it is convenient to define the coordinates of the ellipse parametrically:
\begin{equation}\label{eq:ell_alpha}
    \alpha = a\cos(t),
\end{equation}
\begin{equation}\label{eq:ell_beta}
    \beta = b\sin(t).
\end{equation}
Then an interval along the perimeter of the ellipse, $l$, is given by $dl=\gamma dt$, and the visible portion of the torus surface is defined by $t_1$ and $t_2$, the angles which cause $\mu$ to be largest and smallest for a given $r$ respectively (these are easy to find computationally). Finally, the integral over the torus surface is then,

\begin{equation}\label{eq:u_tor}
\begin{split}
    u(\epsilon, \Omega; r) =& \frac{\xi_{dt}L_{disk}}{(4\pi)^2 c R_{eff}}\delta(\epsilon - 2.7\Theta) \\
    & \times \int_{t_1}^{t_2}\gamma\frac{dt}{x^2}\Big(\frac{R}{R_1}\Big)^{-\zeta}\delta\Big(\mu - \frac{r-\beta}{x}\Big),
\end{split}
\end{equation}
where $\xi_{dt}$, $L_{disk}$, $R_1$, $R_2$, and $R_{eff}$ are defined as before. As in the flat torus model, we weight each $R$ by $R^{-\zeta}$ to account for the radial dependency of the cloud number density. \par
The inner and outer radii of the torus, $R_1$ and $R_2$ respectively, define the semi-major axis of the ellipse, $a=(R_2 - R_1)/2$. Changing $b$ will therefore change the covering fraction of the torus. The covering fraction is defined as follows (see, e.g., \cite{Galanti_2020}):
\begin{equation}\label{eq:f_c}
    f_c = 1 - \frac{\Omega}{4\pi}=\cos(\phi_{max}),
\end{equation}
where $\Omega$ is the solid angle of the visible sky and $\phi_{max}$ is the angle between the jet axis and the tangent of the ellipse that runs through $(r,R)=(0,0)$. This angle can be found by transforming into a coordinate frame in which the cross-section is circular ($\Tilde{\alpha}=\alpha/a$, $\Tilde{\beta}=\beta/b$), where it is simple to find the tangents, and then transforming back. This gives,
\begin{equation}\label{eq:f_c_cos}
    \cos(\phi_{max}) = \frac{b}{\sqrt{b^2 + R_{mid}^2 - a^2}}.
\end{equation}
We use a covering fraction of $f_c=0.6$, which is considered typical \cite{Calderone_2012}. Using $R_1=1.6\times10^{19}$ cm and $R_1=1.2\times10^{20}$ cm as in Ref. \cite{Finke_2016}, this gives a ratio $b/a=0.527$. Table \ref{tab:fields} shows the values of the field parameters used for 3C454.3.

\end{document}